\documentclass[iop]{emulateapj}

\usepackage{natbib}
\usepackage{amsmath}
\usepackage{amsfonts}
\usepackage{hyperref}

\usepackage{color}

\renewcommand{\vec}[1]{\mathbf{#1}}

\begin{document}
\title{Numerical simulations of sunspot decay: On the penumbra -- Evershed flow -- moat flow connection}
\shorttitle{}

\author{M. Rempel\altaffilmark{1}}

\shortauthors{Rempel}

\altaffiltext{1}{High Altitude Observatory,
    NCAR, P.O. Box 3000, Boulder, Colorado 80307, USA}

\email{rempel@ucar.edu}

\begin{abstract}
We present a series of high-resolution sunspot simulations that cover a time span of up to 100 hours. The simulation domain extends about 18 Mm in depth beneath the photosphere and 98 Mm horizontally. We use open boundary conditions that do not maintain the initial field structure against decay driven by convective motions. We consider two setups: A sunspot simulation with penumbra, and a ``naked-spot'' simulation in which we removed the penumbra after 20 hours through a change in the magnetic top boundary condition. While the sunspot has an Evershed outflow of 3-4~km~s$^{-1}$, the naked spot is surrounded by an inflow of 1-2~km~s$^{-1}$ in close proximity. However, both spots are surrounded by an outflow on larger scales with a few 100~m~s$^{-1}$ flow speed in the photosphere. While the sunspot has almost constant magnetic flux content for the simulated time span of 3-4 days, the naked spot decays steadily at a rate of $10^{21}$~Mx~day$^{-1}$.  A region with reduced downflow filling factor, which is more extended for the sunspot, surrounds both spots. The absence of downflows perturbs the upflow/downflow massflux balance and leads to a large-scale radially overturning flow system, the photospheric component of this flow is to the observable moat flow. The reduction of the downflow filling factor also inhibits submergence of magnetic field in the proximity of the spots, which stabilizes them against decay. While this effect is present for both spots, it is more pronounced for the sunspot and explains the almost stationary magnetic flux content. 
\end{abstract}


\received{}
\accepted{}

\maketitle
\section{Introduction}
\label{sect:intro}
Large scale flows are an integral part of sunspot structure. The most prominent flow is the Evershed flow in the sunspot penumbra, which was discovered by \citet{Evershed:1909}. This flow component has been successfully modeled in recent magnetoconvection simulations such as \citet{Heinemann:etal:2007,Scharmer:etal:2008,Rempel:etal:2009,Rempel:etal:Science,Kitiashvili:etal:2009,Rempel:2011,Rempel:2012:penumbra} as consequence of overturning convection in the presence of an inclined magnetic field. On even larger scales sunspots are surrounded by moat flows, which were first found through tracking of magnetic features \citep{Sheeley:1969,Vrabec:1971,Harvey:Harvey:1973}, Doppler measurements \citep{Sheeley:1972} \citep[see][for a review of these early discoveries]{Vrabec:1974}, and later helioseismic measurements \citep{Gizon:etal:2000}. Moat flows have typically flow velocities of a few $100$ m s$^{-1}$ and extend up to 2 sunspot radii \citep{Brickhouse:Labonte:1988}. Helioseismic inversions by \citet{Featherstone:etal:2011:JPC} find outflows around sunspots with an amplitude of $100$~m~s$^{-1}$ to a depth of about $7$~Mm, below that the amplitude drops significantly.

It is currently an open debate if moat flow and Evershed flow (or at least the presence of a penumbra) are related.  Support for a connection (or at least some common cause) comes from observations of spots with asymmetric penumbrae, in which moat flows are only present in radial extension of penumbral filaments and absent in the perpendicular directions as well as spot segments with no penumbra at all  \citep{VDominguez:etal:2007,VDominguez:etal:2008}. Links are also suggested by observations that show moving magnetic features traveling from within the penumbra into the moat region \citep{SDalda:MPillet:2005}. In contrast to this a recent study by \citet{Loehner-Boettcher:2013:moat} observing a total of 31 sunspots did not find a convincing correlation of the moat flow with sunspot size or Evershed flow speed and concluded that the moat flow is of independent origin.

Additional insight on the penumbra - Evershed flow - moat flow connection can be gained from comparing the flow properties in the vicinity of pores and naked sunspots with those of fully developed sunspots. The dominant flow in the absence of a penumbra is an inflow \citep{Sobotka:etal:1999,VDominguez:etal:2010,SainzDalda:2012:naked}. Interestingly, diverging flows are found further out, but they are not typically classified as moat flows \citep{VDominguez:etal:2010}. The comparison of pores and naked spots with sunspots is non trivial since pores and naked spots have typically a smaller size and also a different evolutionary history, i.e. it is not obvious that potential differences can be solely attributed to the absence of a penumbra.

Models of moat flows typically link this flow to deeper seated flow systems around sunspots. \citet{Meyer:etal:1974} suggested that sunspots initially form in a super-granular vertex. Due to heatflux blockage the converging flow toward the spot turns eventually into an outflow. They suggested that the outflow turns into an inflow in more than $10$~Mm depth and that such converging flows play a central role in confining the magnetic field of a sunspot. \citet{Parker:1979:cluster} suggested that sunspots are confined by converging flows in the uppermost few Mm of the convection zone, which would lead to a flow opposite to the moat flow. It was suggested later by \citet{Hurlburt:Rucklidge:2000} that the converging ``collar flow'' might be hidden beneath the penumbra with diverging Evershed flow. \citet{Nye:1988:moat} modeled flows as a consequence of heatflux blockage by umbra and penumbra and found that the depth extent of the penumbra plays a central role in determining the extent of the resulting outflow. 

Numerical simulations of sunspots including the moat region are challenging. Resolving a penumbra requires a minimum resolution, capturing the moat region requires large domains and also substantially longer time scales -- in other words -- computationally expensive setups. 

Early work used idealized setups without a realistic photosphere and focused on flow systems around magnetic flux concentrations in axisymmetric setups \citep{Hurlburt:Rucklidge:2000,Botha:etal:2006}. Here it was found that stable flux concentrations are surrounded by a converging ``collar flow'' at the top of the domain, similar to the flows found around pores. It was argued that such flow also exists beneath a sunspot penumbra with outward directed Evershed flow to maintain stability of the spot \citep[see also][]{Parker:1979:cluster}. This work was generalized to non-axisymmetric simulations by \citet{Botha:etal:2011}. While similar flow patterns were found on average, the non-axisymmetric flow structure allowed here magnetic flux to escape from the central flux concentration. 

Numerical simulations with a realistic photosphere in domains large enough to capture a moat region were presented by \citet{Rempel:2011:moat}. In particular that investigation focused on a comparison of numerical sunspot models with and without penumbra. It was found that spots without penumbra (i.e. pores or naked spots) do have a converging flows in their vicinity similar to observational findings. However, further out diverging flows are present with properties similar to those found around sunspots. Due to computational constraints the setups used in \citet{Rempel:2011:moat} were not fully comparable, i.e. the sunspot and naked spots had different flux content and were simulated in domains with different extent as well as resolution. Here we will expand the work of \citet{Rempel:2011:moat} and present a series of numerical simulations that compare again a sunspot with a naked spot. The numerical setups are identical in both cases, with exception of the magnetic top boundary condition through which we control the extent of a penumbra. At least initially sunspot and naked spot have the identical magnetic flux. We do use in our setups boundary conditions that do not inhibit spot decay and we do run the simulations for up to $100$ hours of simulated time in order to be able to address also the role of a penumbra and associated flows for sunspot decay.

\section{Simulation setup}
\label{Sec:setup}

We present here sunspot simulations that are based on the \textit{MURaM} radiative MHD code \citep{Voegler:etal:2005,Rempel:etal:2009,Rempel:2014:SSD}. In particular we use here the version of the code that has been recently described in detail by \citet{Rempel:2014:SSD} and used therein for photospheric small-scale dynamo simulations. The MURaM code solves the MHD equations using a fourth order accurate finite
difference scheme combined with short characteristics radiative transfer \citep{Voegler:etal:2005}. For numerical stability artificial diffusivities are added using a slope-limited diffusion scheme as detailed in \citet{Rempel:2014:SSD}. For the sunspot simulations presented here a special treatment of low-$\beta$ regions ($\beta=8\pi p_{\rm gas}/B^2$) is essential. As described in \citet{Rempel:2014:SSD} we use an energy equation that considers only internal and kinetic energy and separates out the magnetic energy in order to avoid instabilities in low-$\beta$ regions and we artificially reduce the Lorentz force in order to limit the Alfv{\'e}n velocity \citep{Rempel:etal:2009,Rempel:2014:SSD}. In the simulation presented here the Alfv{\'e}n velocity is limited at $75$~km/s, which happens mostly in the area above the sunspot umbra and penumbra in height of more than $200$~km above the $\tau=1$ surface. In these regions the temperature remains low near $4000-5000$~K, i.e. the speed of sound is near $6$~km/s. This ensures that the limited Lorentz force remains the dominant force, i.e. the magnetic field has to remain close to a force-free field configuration.

One of the main differences to the code described in \citet{Rempel:2014:SSD} is as follows. For the sunspot simulations presented here we found a drift of the total magnetic flux in the simulation domain on the order of $10^{-5}$ G per time step. While this drift has not been a significant problem in previous simulations that addressed only the shorter term evolution, the simulations presented here are evolved for up to $100$ hours, requiring up to $800,000$ time steps. Over the duration of the simulation this drift would have added up to about $20$ \% of the flux content of the sunspots. In a first series of simulations we corrected the drift by adding a correction term to the vertical component of the induction equation:
\begin{equation}
	\frac{\partial B_z}{\partial t}=[\ldots] +\frac{1}{\tau}(B_0-\langle B_z\rangle)\,.
	\label{Eq:flux_corr}
\end{equation}
Here $\langle B_z\rangle$ denotes the horizontally averaged vertical magnetic field, $B_0=\Phi/A$ the average field strength corresponding to the netflux $\Phi$ of the simulation domain and $\tau$ is a relaxation time scale chosen to be 100 time steps (45 seconds). While this prevented the drift, we remained concerned that the systematic differences we see between our sunspot and naked spot simulation described below are at least partially caused by this issue since the applied corrections had opposing signs (a positive drift in the case of the sunspot and a negative drift for the naked spot). In order to verify that our method of removing the drift does not impact the simulation results, we conducted two additional experiments with modified numerical diffusivities. It turns out that the numerical diffusive flux of $B_z$ in the $z$-direction leads to a systematic transport of magnetic flux across the top boundary and that the drift problem can be alleviated by setting this flux to zero. We went even one step further by setting throughout the simulation domain the diffusive fluxes of $B_x$ in the $x$, $B_y$ in the $y$ and $B_z$ in the $z$-direction to zero, which does not impact numerical stability, but reduces the overall $\vec{\nabla}\cdot\vec{B}$ error produced by the scheme (in addition the numerical scheme is also more consistent with the fundamental property of the induction equation that the time derivative of $\vec{B}$ is given by a curl or alternatively the divergence of an anti-symmetric flux tensor, i.e. fluxes in the direction of the field do not exist).  As we describe in Section \ref{sec:time_evol}, both approaches lead to similar results and we are confident that the drift problem does not impact any of the results presented here. The remaining $\vec{\nabla}\cdot\vec{B}$ error of the numerical scheme is corrected by using the hyperbolic $\vec{\nabla}\cdot\vec{B}$  cleaning method of \citet{Dedner:etal:2002:divB}.

We use here a bottom boundary condition that imposes a symmetric massflux and magnetic field across the boundary, i.e. it allows for the presence of vertical and horizontal magnetic field and flow components (boundary ``OSb'' described in \citet{Rempel:2014:SSD}). Similar to previous work, horizontal boundary conditions are periodic and at the top we use the same boundary condition that was used in \citet{Rempel:2012:penumbra}. The top boundary is closed for flows and magnetic field components in the boundary layer ``ghost cells'' are derived from the vertical magnetic field at the top of the domain through a convolution with kernels that are described in the Appendix B of \citet{Rempel:2012:penumbra}. The kernels impose a relationship between vertical and horizontal field components as well as vertical decay rate of field components, but allow otherwise for a free evolution of the field, i.e. a potential decay of the sunspot is not inhibited by the boundary condition.

\begin{figure*}
  \centering 
  \resizebox{0.75\hsize}{!}{\includegraphics{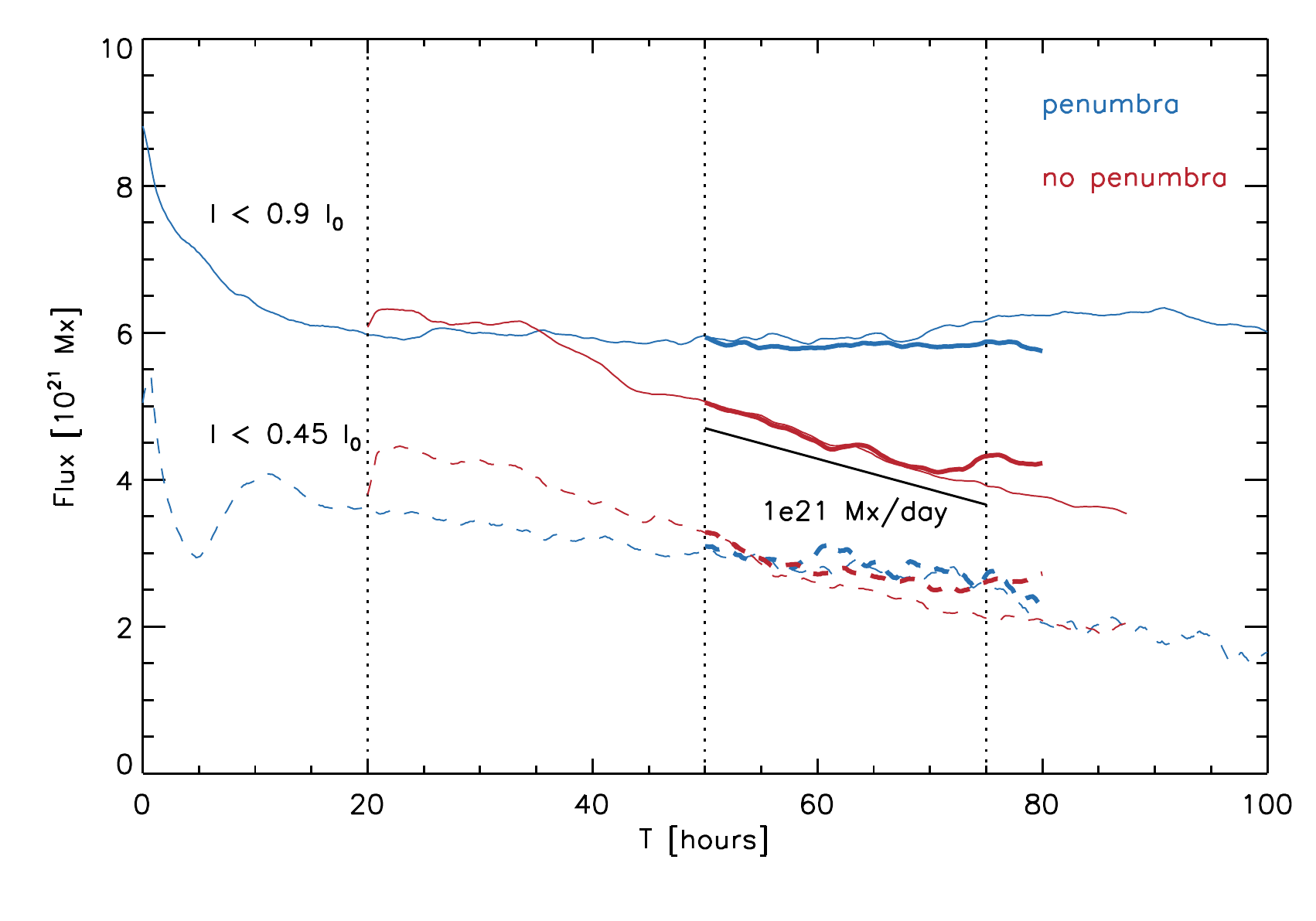}}
  \caption{Time evolution of the magnetic flux in the photosphere in regions with $\bar{I}<0.45 I_{\odot}$ (dashed) and $\bar{I}<0.9 I_{\odot}$ (solid). Blue lines correspond to the sunspot with penumbra, red lines to the "naked spot'' without penumbra. In addition the thick lines correspond to the simulations with modified numerical diffusivity as described in the text.}
  \label{fig:01}
\end{figure*}

The detailed simulation setup is a compromise between resolution and domain size: 1. Sufficient resolution to capture penumbral fine structure, 2. A large enough domain to capture a sunspot including the moat region, and 3. A domain size not too large in order to allow for simulations covering several days of solar time. \citet{Rempel:2012:penumbra} found that a grid spacing of at least $48$~km horizontally and $24$~km vertically is required to capture penumbral fine structure and the related Evershed flow. In order to capture the moat region of a sunspot, which typically extends to about 2 spot radii, we have chosen an overall domain extent of $98.304\times 98.304\times 18.432$~Mm$^3$ leading to a size of $2048\times 2048\times 768$ grid cells. Similar to previous setups the photosphere is located about $700$~km beneath the top boundary, leading to an almost $18$~Mm deep convection zone with a density contrast of close to $2\times 10^4$ from the bottom of the simulation domain into the photosphere (the whole convection zone would be a contrast of about $10^6$). The average density contrast from the photosphere to the top boundary of the simulation domain is about a factor of $10^3$ in addition.

The initial state is a relaxed small scale-dynamo simulation similar to those discussed in \citet{Rempel:2014:SSD}. While a grid spacing of $48$~km horizontally and $24$~km vertically is at best marginal for resolving an SSD in the photosphere (\citet{Rempel:2014:SSD} found that about $50$\% of the magnetic energy resides on scales smaller than $100$~km), we do find an unsigned vertical flux of about $50$~G at $\tau=1$, not too different from the expected value for the quiet Sun (about $60$~G, see \citet{Rempel:2014:SSD} for a detailed discussion). We initialize the sunspot simulation by inserting an axisymmetric self-similar magnetic field structure into the the domain, following the description found in the Appendix A of \citet{Rempel:2012:penumbra}. The total initial flux is $9\times 10^{21}$ Mx, the field strength at the bottom of the domain is around $20$~kG, the field strength at the top of the domain around $3$~kG (the detailed parameters as defined in the Appendix A of  \citet{Rempel:2012:penumbra} are $B_0=20.25$~kG, $R_0=4$~Mm, and $z_0=13.38$~Mm).

For the first $2\times 10^4$ seconds of the time evolution we impose a damping term in the momentum equation within a cylinder of $15$~Mm radius around the inserted flux concentration. The initial damping time scale is $100$ seconds and increases exponentially with an e-folding time scale of $3000$ seconds. During this time we also close the bottom boundary for vertical flows (asymmetric vertical massflux across boundary) within the cylinder. The purpose of this initial phase is to allow the inserted magnetic field structure to settle into a force equilibrium and to setup a convective flow field in the proximity of the spot that is consistent with the presence of the spot. After this initialization phase we switch back to the open boundary condition described above for the whole computational domain. This boundary condition does not further constrain the evolution of the spot, in particular it does not inhibit a possible decay of the spot. 

Sunspot decay is likely a combination of several processes operating at different depth beneath the photoshere. In our numerical setup we can only capture processes in the uppermost 10-20 Mm of the convection zone, in particular we are considering (1) turbulent erosion and submergence due to near surface convection (motions on scales smaller than the spot size) and (2) turbulent erosion of the ``footpoint'' of a sunspot due to deeper seated convection (motions on scales comparable or larger than the spot size). Our setup is an attempt to minimize the influence from the latter and focus on the former. Process (1) is ubiquitous and should equally affect all sunspots, while process (2) is likely highly dependent on the details of individual spots and their formation history. In numerical simulations an additional contribution could come from numerical diffusivities. We will quantify their contribution in the end of Section \ref{sec:phot_app}.

In addition our aim is not to study spot decay in absolute terms. Instead we will compare spot decay simulations and analyze in particular differences that are introduced by the presence or absence of a penumbra. To this end we compare two setups. The ``sunspot simulation'' uses a magnetic top boundary condition that imposes a sufficiently horizontal field to maintain a penumbra. We follow here the approach detailed in Appendix B of \citet{Rempel:2012:penumbra} and use a top boundary with a parameter of $\alpha=2$ (horizontal magnetic field components are increased by a factor of $2$ compared to a potential field). \citet{Rempel:2012:penumbra} explored values of $\alpha$ from $1.5$ to $2.5$, which all lead to penumbrae with strong Evershed flows. We use this setup to initialize our run as described above and evolve the simulation for $100$~hours of solar time. The ``naked spot simulation'' is restarted from a snapshot at $20$~hours from the sunspot simulation and uses a top boundary with the setting of $\alpha=1$ (potential field). The change of the magnetic top boundary condition leads to the disappearance of the previously developed penumbra within about $30$~minutes  (we note that switching back to the $\alpha=2$ boundary condition does lead to the re-appearance of the penumbra on a similar time scale). We evolve this simulation for $67.5$ hours. Most of the following analysis will be based on snapshots starting at $t=50$~hours, i.e. $50$ hours after initialization for the sunspot setup and $30$ hours after the boundary condition change for the naked spot setup.  Both boundary conditions satisfy the $\vec{\nabla}\cdot\vec{B}$  constraint, however only the  $\alpha=1$  (potential field) boundary condition is current free. In the case of the $\alpha=2$ boundary condition we have a ring current flowing outside the simulation domain that adds a magnetic field opposing the sunspot field, leading to a solution with an overall more strongly inclined magnetic field near the top boundary. This ring current is only present outside the simulation domain, the magnetic field in the upper layers of the simulation domain has to remain close to force-free, with no such current present owing to the low $\beta$ condition realized there.

All simulations presented here are computed with gray radiative transfer. When we refer in the following discussion to intensity and optical depth surfaces, we consider always quantities that are computed with respect to the gray mean opacity and source function.

\section{Results}

\begin{figure*}
  \centering 
  \resizebox{0.75\hsize}{!}{\includegraphics{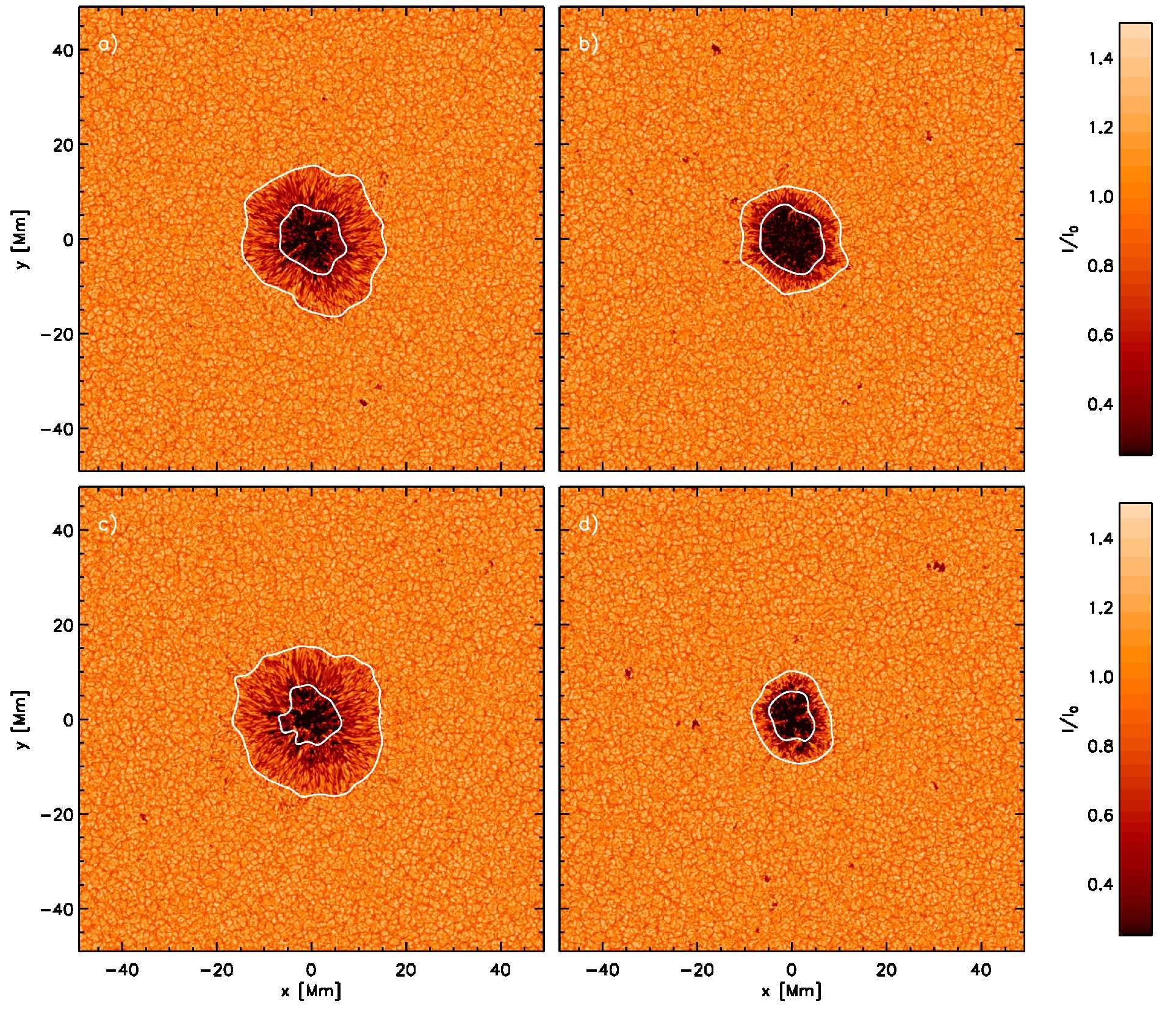}}
  \caption{Comparison of sunspot (left) and ``naked spot'' (right) through intensity images. Panels a) and b) correspond to snapshots at $t=25$~hours, panels c) and d) at $t=75$~hours. Contour lines indicate the regions $\bar{I}<0.45 I_{\odot}$ and $\bar{I}<0.9 I_{\odot}$. For the sunspot (left panels) the umbra area decreases by $25\%$, while the total spot area increases by $11\%$. In the case of the naked spot, the umbra  (total spot) areas decrease by $49\%$ ($36\%$). The relative fraction of umbra area decreases for the sunspot from $20\%$ to $14\%$ and in the case of the
naked spot from $38\%$ to $30\%$.}  
  \label{fig:02}
\end{figure*}

\subsection{Time evolution of magnetic flux}
\label{sec:time_evol}

Figure \ref{fig:01} presents the time evolution of magnetic flux for the sunspot/naked spot simulations discussed in this paper. Thin lines correspond to the simulations that were computed with the flux correction Eq. \ref{Eq:flux_corr}, solid lines correspond to simulations that were computed with modified numerical diffusivities as described in Section \ref{Sec:setup}. We describe the former first. We compute the flux content of areas in the photosphere by considering masks for the regions with $\bar{I}<0.45 I_{\odot}$ (``umbra'', dashed lines) and $\bar{I}<0.9 I_{\odot}$ ``umbra+penumbra'', solid lines). Here $\bar{I}$ denotes an intensity that is smoothed through a convolution with a Gaussian with FWHM of $3$~Mm. The resulting masks are shown in Figure \ref{fig:02} for a snapshots at $t=25$ and $75$~hours. The flux is computed using $B_z$ on a constant height surface that corresponds to the average $\tau=1$ level in the plage region surrounding the spots. We found that using $B_z$ on a constant $\tau$ surface (as it is typical in observations) has the tendency to overestimate the flux content of a spot by $10-15\%$. The blue lines in Figure \ref{fig:01} correspond to the sunspot with penumbra. It takes about $15$~hours from the initialization of the simulation until the spot is settled. While we start with $9\times 10^{21}$~Mx flux, about $6\times 10^{21}$~Mx are left in the spot after $15$~hours. While we do use boundary conditions that allow for spot decay, we find an almost stationary solution for the magnetic flux contained in the region with $\bar{I}<0.9 I_{\odot}$  until $t=100$~hours, where we stopped the simulation.  From $t=70$ to $90$~hours we find  a gradual increase of flux, which is related to the re-emergence of magnetic field in the proximity of the spot that was submerged earlier during the simulation. In contrast to the region $\bar{I}<0.9 I_{\odot}$, the ``umbra'' ($\bar{I}<0.45 I_{\odot}$) of the sunspot shows a steady decay after $t=15$~hours with an average rate of about $6\cdot 10^{20}$~Mx~day$^{-1}$. Since the flux content of the region $\bar{I}<0.9 I_{\odot}$ is almost steady, this implies a steady increase of the flux content (and area) of the penumbra (see also Figure \ref{fig:02}). This behavior is possibly inconsistent with the observed decay of sunspot, where the penumbra area decays more quickly than the umbra area \citep{Deng:etal:2007:flow_decaying_sunspot}, but since we did not run our simulations for long enough we cannot address the final stages of decay here. The relative umbra area in the range of $14-20\%$  is similar to observed values \citep[see, e.g.,][]{Solanki:2003}.  

At $t=20$~hours we started from our sunspot simulation a new simulation that differs in terms of the top boundary condition. We switched to a potential field boundary, which effectively decreases the strength of the horizontal magnetic field at the top boundary by about a factor of $2$. This change leads to a disappearance of the penumbra within about $30$ minutes and we refer to this simulation in this paper as ``naked spot''. The corresponding flux evolution is shown in Figure \ref{fig:01} in red color.  Initially the flux in both umbra and total spot area increases since the spot becomes more concentrated and stronger in response to the less inclined field at the top boundary. For about $15$~hours the total flux of the spot does not show any signs of enhanced decay compared to the sunspot. Starting from about $t=35$~hours the total flux content of the spot declines at an average rate of about $1.1\times 10^{21}$~Mx~day$^{-1}$. The flux of the umbra decays at an average rate of about $9\times 10^{20}$~Mx~day$^{-1}$ and approaches toward the end of this simulation $t=87.5$~hours the flux found in the umbra of the spot with penumbra.

\begin{figure*}
  \centering 
  \resizebox{0.75\hsize}{!}{\includegraphics{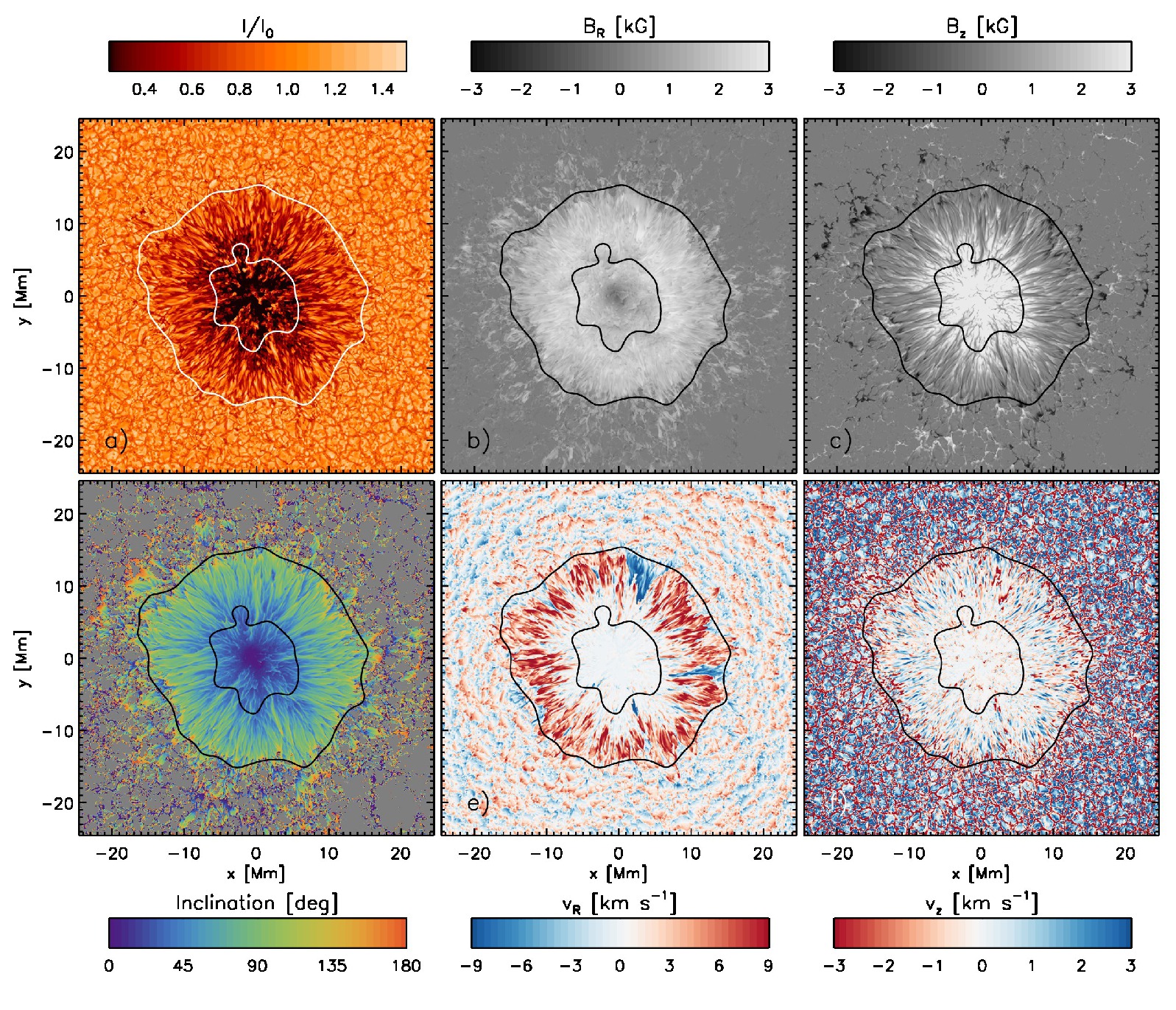}}
  \caption{Photospheric ($\tau=1$) fine structure of sunspot (the panels show only the innermost $49.152\times 49.152$~Mm$^2$ of the simulation domain). Presented are a) intensity, b) radial field strength, c) vertical field strength, d) field inclination with respect to vertical (regions with $\vert B\vert<200 G$ are masked with grey color for enhanced clarity), e) radial flow velocity, and f) vertical flow velocity. We show a snapshot at $t=70$~hours. An animation covering $25$ hours is provided with the online material.}
  \label{fig:03}
\end{figure*}

We conducted two additional experiments that were started from snapshots at $t=50$~hours and ran until $t=80$~hours and use the modified numerical diffusivities as described in Section \ref{Sec:setup}. They
are indicated through thick lines. The basic result that the sunspot is almost stationary, while the naked spot decays at a rate of about $10^{21}$~Mx~day$^{-1}$ is confirmed, although differences exist in detail, which is expected due to the non-linear nature of this problem. For example we see an increase of the naked spot flux starting from about $t=73$~hours, which is related to re-emerging magnetic flux. This flux emergence event is captured in the animation we provide for Figure \ref{fig:04} in the lower left corner of the naked spot. In our original runs this happend in the case of the sunspot instead. The following analysis is based on data from these new simulations.

\subsection{Photospheric appearance of spots}
\label{sec:phot_app}

\begin{figure*}
  \centering 
  \resizebox{0.75\hsize}{!}{\includegraphics{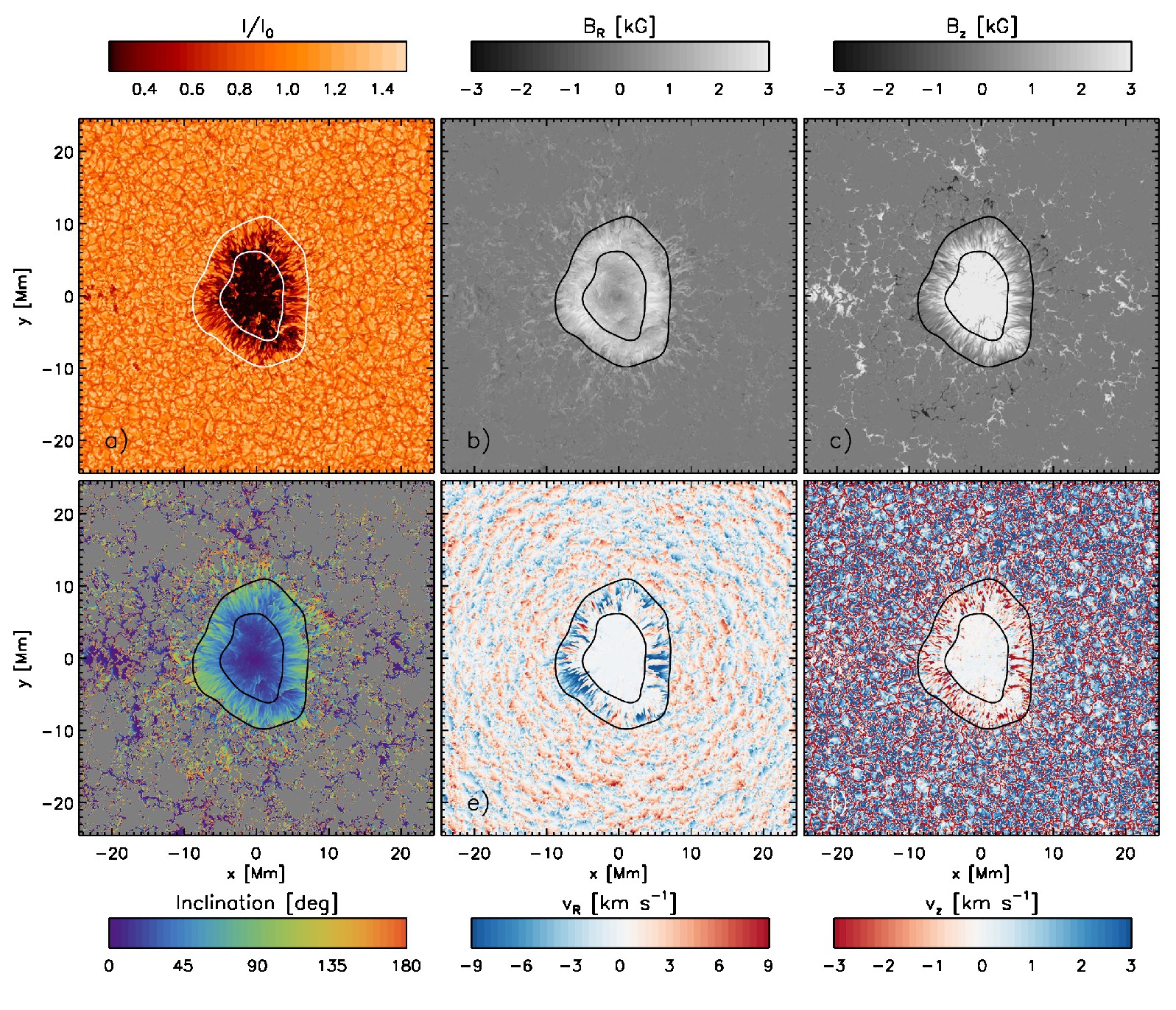}}
  \caption{Photospheric ($\tau=1$) fine structure of naked spot (the panels show only the innermost $49.152\times 49.152$~Mm$^2$ of the simulation domain). All quantities are the same as shown in Figure \ref{fig:03}. We show a snapshot at $t=70$~hours. An animation covering $25$ hours is provided with the online material.}
  \label{fig:04}
\end{figure*}

\begin{figure*}
  \centering 
  \resizebox{0.7\hsize}{!}{\includegraphics{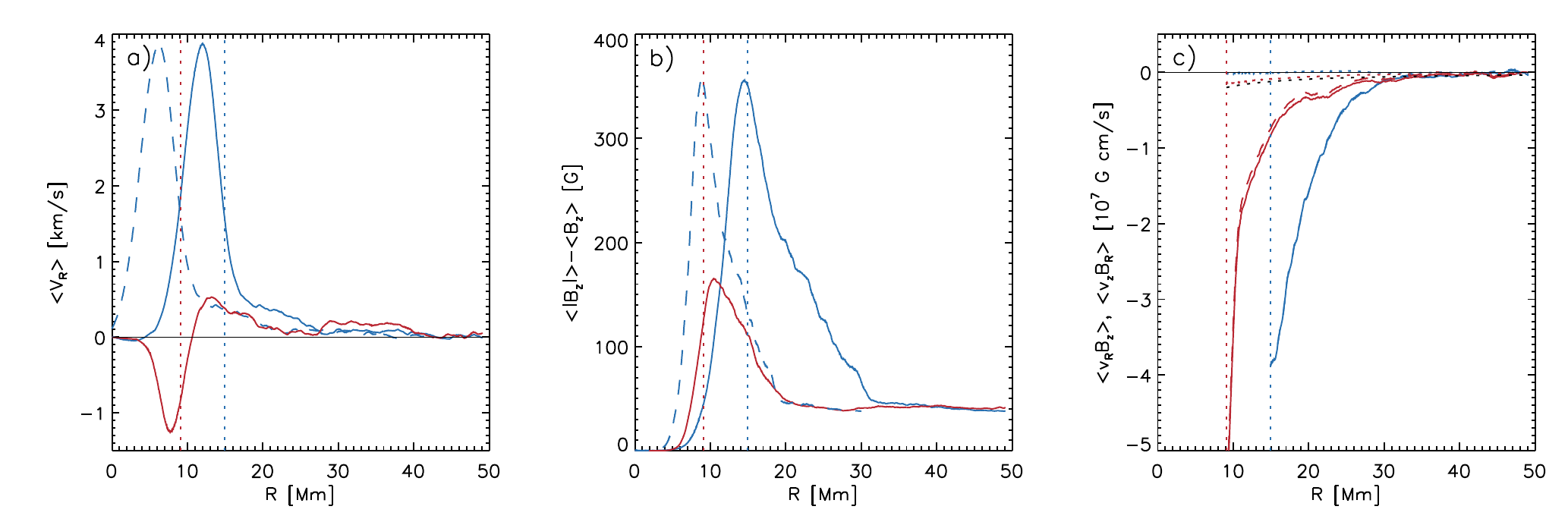}}
  \caption{Azimuthal and temporal averages (from 50 to 75 hours) of a) $\overline{v_R}$, b) $\overline{\vert B_z\vert}-\overline{B_z}$, and c) $\overline{v_z B_R}$ (solid) and  $\overline{v_R B_z}$ (dashed). Blue colors correspond to the sunspot, red colors to the naked spot. Vertical dotted lines indicate the radius for both spots at which $\bar{I}$ reaches $90$ \% of the average intensity outside the spots. In panels a) the blue dashed line indicates the flow profile of the sunspot shifted by $5.8$~Mm (difference between spot radii) to the left. In panel b) the blue dashed line indicates the magnetic flux profile with the radial distance rescaled by a factor of $0.61$ (ratio of spot radii). In panel c) dotted lines indicate $\overline{v_z B_R} - \overline{v_R B_z}$. The black dotted line indicates values that would correspond to a spot decay of $10^{21}$ Mx~day$^{-1}$. The quantities in panels a) and b) are evaluated on the warped $\tau=1$ surface, while quantities in panel c) are computed on a constant height surface corresponding to $\tau=1$ outside the spots. In panel c) we do not display values inside the spots.}
  \label{fig:05}
\end{figure*}

\begin{figure*}
  \centering 
  \resizebox{0.75\hsize}{!}{\includegraphics{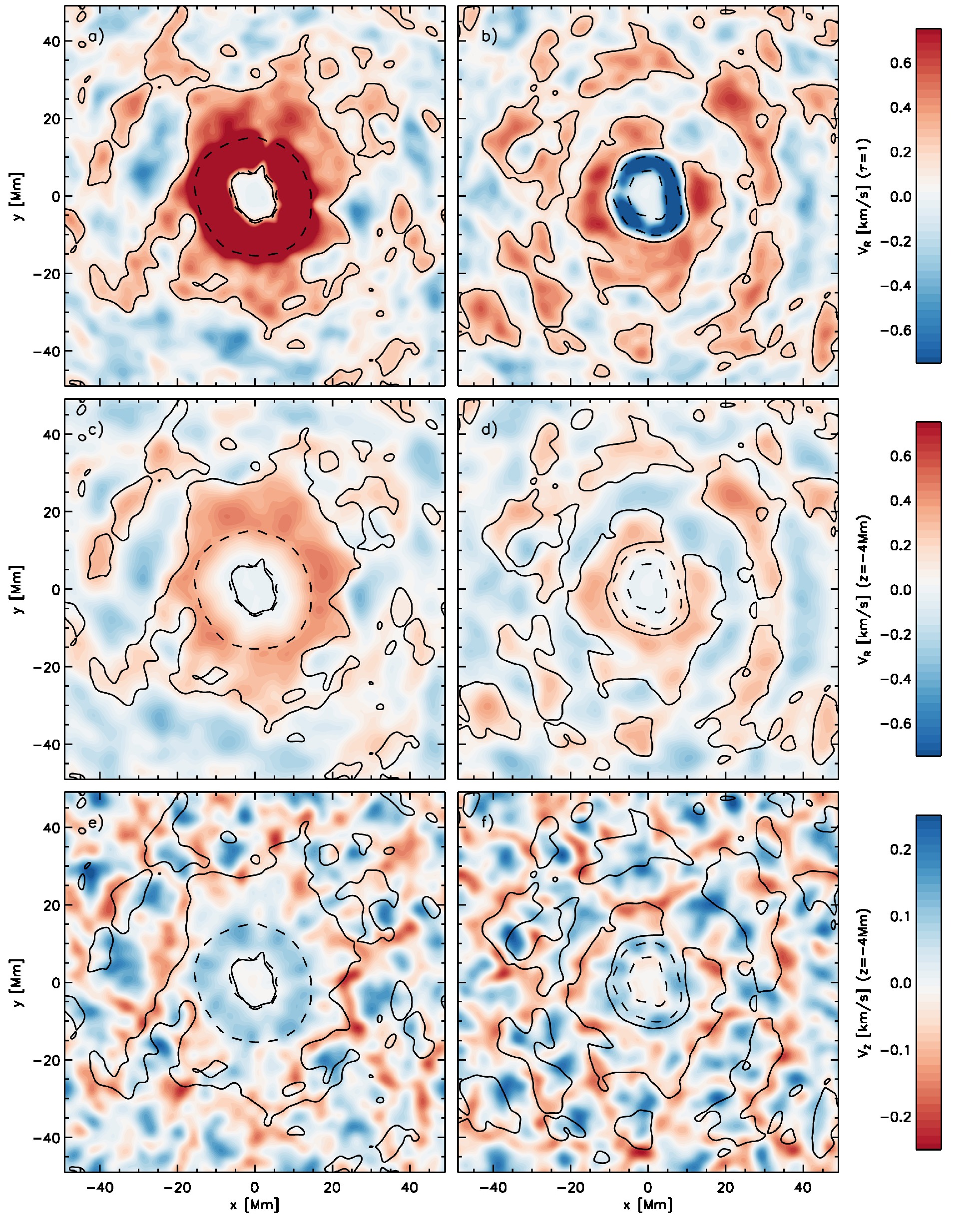}}
  \caption{Comparison of sunspot (left) and naked spot (right). Panels a) and b) show the radial flow velocity with respect to the center of the spots on the $\tau=1$ level, panels c) and d) show $v_r$ in $4$~Mm depth, and panels (e) and f) show $v_z$ in $4$~Mm depth (upflows are positive). In order to emphasize long-lived large-scale flow components we present $25$~hour time averaged (from 50 to 75 hours) and horizontally (Gaussian with a FWHM of $3$~Mm) smoothed quantities. Solid contours enclose regions with radial outflows in the photosphere of more than $200$~m~s$^{-1}$. Dashed contours indicate the regions with $\bar{I}<0.45 I_{\odot}$ and $\bar{I}<0.9 I_{\odot}$.}
  \label{fig:06}
\end{figure*}

\begin{figure*}
  \centering 
  \resizebox{0.75\hsize}{!}{\includegraphics{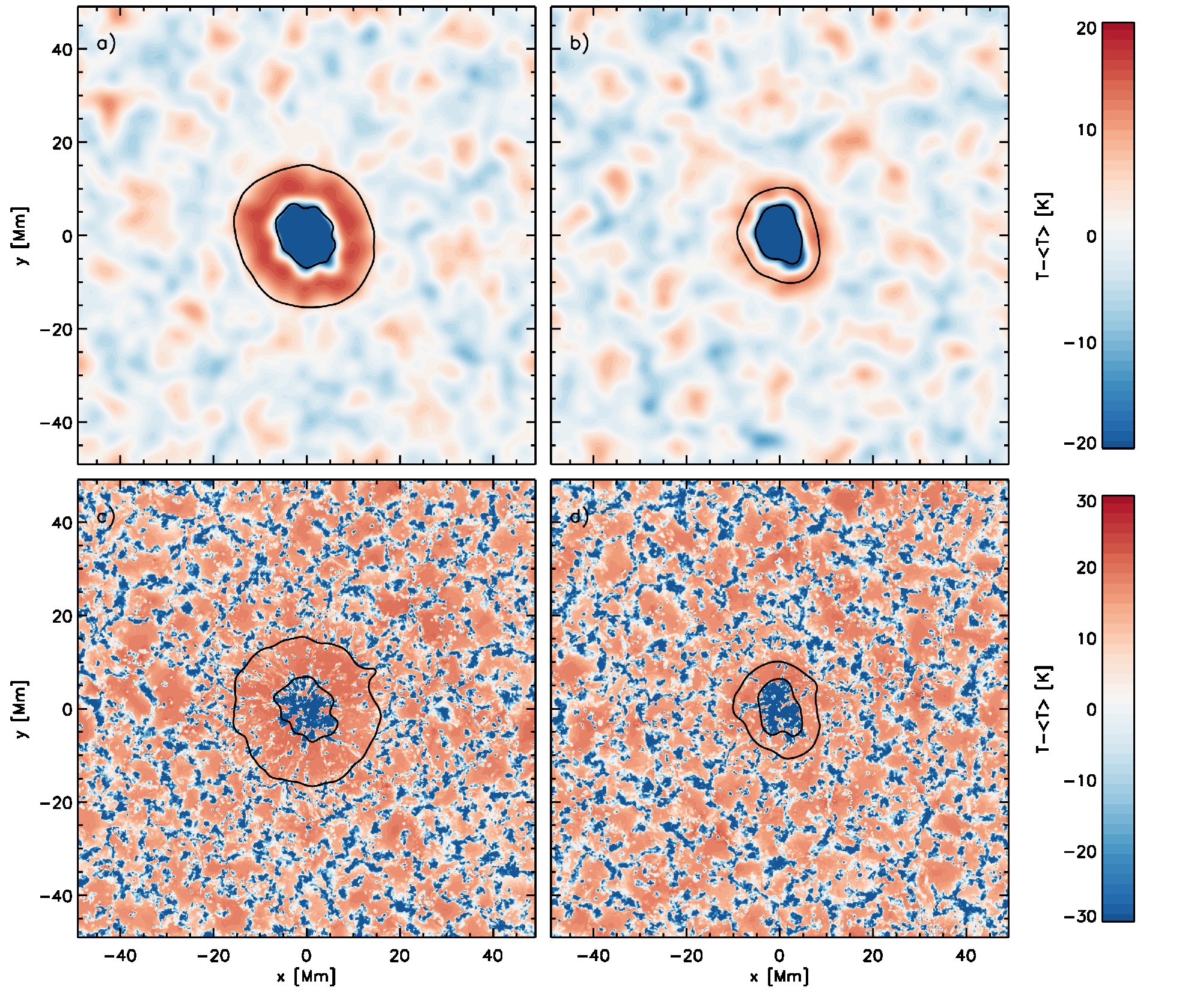}}
  \caption{Comparison of subsurface temperature structure for both spots in $4$~Mm depth. Panels a) and b) show the $25$~hour time averaged (from 50 to 75 hours) and spatially (Gaussian with a FWHM of $3$~Mm) smoothed temperature perturbation, panels c) and d) show the temperature perturbation for a 15 minute averaged snapshot (at $t=62.5$~hours). Solid contours indicate the regions with $\bar{I}<0.45 I_{\odot}$ and $\bar{I}<0.9 I_{\odot}$ for both spots.}
  \label{fig:07}
\end{figure*}

Figures \ref{fig:03} and \ref{fig:04} summarize the photospheric ($\tau=1$) appearance of the sunspot and naked spot, respectively. In both cases we show a snapshot at $t=70$~hours. The quantities shown are identical to those discussed in \citet[][Figure 6] {Rempel:2012:penumbra} for a sunspot simulation with 3 times higher horizontal and two times higher vertical resolution. Despite the lower resolution we find for the sunspot case (Figure \ref{fig:03}) the basic building blocks of penumbral fine structure: radially aligned filaments with close to horizontal field (panel d) due to the strong reduction of the vertical field component (panel c), while the horizontal field remains strong (panel b).  Along these filaments we find fast radial outflows (panel e) reaching peak flow speeds of 10 km~s$^{-1}$. The azimuthally averaged flow reaches a peak flow speed of about 4 km~s$^{-1}$ (see Figure \ref{fig:05}). In addition to these strong outflow regions, we do find also patches in the penumbra that do have even at $\tau=1$ an inverse Evershed flow. These region are transient and typically last a few hours, before they return into outflow regions again (see the animation of Figure \ref{fig:03} provided with the online material). We do not further analyze this feature here, but point out that a similar behavior (anomalous flow in photosphere) has been found in observations by \citet{Kleint:Saina-Dalda:2013:unusual-filaments,Louis:2014:anomalous_flow_penumbra}. 

In the naked spot simulation (Figure \ref{fig:04}) we do not find an extended penumbra and the spot is on average surrounded by a converging flow with a few km~s$^{-1}$ amplitude. Nonetheless there are a few locations in which we find intermittent filaments that host an outflow. Further away from the spots we do find in both cases a region with an enhanced amount of mixed polarity flux in the photosphere (panels c). The animations provided for
Figures \ref{fig:03} and \ref{fig:04} indicate a diverging flow away from the approximate spot center, which is presented in Figure \ref{fig:06}.  

Figure \ref{fig:05}a) presents for both spots the azimuthally and temporally averaged radial flow velocity $\overline{v_R}$ with respect to approximate center of the spots.  For the sunspot we find the Evershed flow peaking at about 4 km~s$^{-1}$ at $R=12$ Mm. Outside the sunspot we find a radial outflow which drops below $200$ m~s$^{-1}$ at  $R_{\rm flow}=25$ Mm (``moat region''). While the naked spot shows an inflow of up to $1.2$ km~s$^{-1}$ at its periphery, it is also surrounded further out by an outflow dropping below $200$ m~s$^{-1}$ at $R_{\rm flow}=19$ Mm. The radii of sunspot and naked spot are $R_{\rm spot}=14.9$ and $9.1$ Mm, respectively. The moat flow extends to about $1.7 R_{\rm spot}$ for the sunspot and $2.1 R_{\rm spot}$ for the naked spot, indicating that there is not simply a proportional relation between $R_{\rm flow}$ and $R_{\rm spot}$. The blue dotted lines indicates the flow profile of the sunspot shifted $5.8$~Mm (difference in spot radii) to the left, which leads to a good match with the flow profile of the naked spot. This indicates a relation 
\begin{equation}
	R_{\rm flow}\approx R_{\rm spot}+R_0\;,
	\label{EQ:Rmoat}
\end{equation}
where $R_0\sim 10$ Mm in the simulations presented here (using the $200$ m~s$^{-1}$ threshold for defining the outer boundary). There is another outflow patch with $200-300$ m~s$^{-1}$ found between $R=30$ and $40$ Mm, which is related to a ring-like arrangement of convection cells that becomes more evident in Figures \ref{fig:06} and \ref{fig:09}. While not completely absent, this feature is less pronounced for the sunspot. 

Figure \ref{fig:05}b) characterizes the amount of mixed polarity magnetic flux through the quantity $\overline{\vert B_z\vert-B_z}$. The asymptotic value of about $40-50$ G for $R>30$ Mm is a consequence of a small-scale dynamo operating throughout the simulation domain. If we use a threshold of $60$~G to define the extent of the enhanced mixed polarity flux region we find for the sunspot $R_{\rm flux}=30.3$~Mm and for the naked spot $R_{\rm}=18.4$~Mm, indicating a relation
\begin{equation}
	R_{\rm flux}\approx 2 R_{\rm spot}\;.
	\label{EQ:Rflux}
\end{equation}
The blue dashed line in Figure \ref{fig:05}b) indicates this relationship (the radial axis is rescaled to the radius of the naked spot). The amount of mixed polarity flux is about 2 times larger for the sunspot, which has also a 2 times stronger magnetic canopy field overlying the photosphere due to the different magnetic top boundary conditions used here.

The time evolution of the magnetic flux contained within a radius $R$ is given by the expression \citep[see also][]{Cheung:etal:2010,Rempel:Cheung:2014:fem}:
\begin{equation}
\dot{\Phi} = 2\pi R\left( \overline{v_z B_R}- \overline{v_R B_z}\right)\;.
\label{eq:flux_evol}
\end{equation}
In Figure \ref{fig:05}c) we present the contributions from the terms $\overline{v_z B_R}$ and $\overline{v_R B_z}$. Possible modes of spot decay (in our setup the spot has positive polarity) are submergence of horizontal (radial) magnetic field ($\overline{v_z B_R}<0$) or radial outward transport of vertical flux elements with the polarity of the spot ($\overline{v_R B_z}>0$). The situation we find here is however quite different. Both terms have a large amplitude and negative sign and cancel each other almost perfectly (as a consequence only the solid blue line is visible in Figure \ref{fig:05}c). We do find a strong term $\overline{v_z B_R}<0$, which would lead if unbalanced to a spot decay at a rate exceeding $10^{22}$ Mx~day$^{-1}$ for both spots. This term is however compensated to a large degree by a term $\overline{v_R B_z}<0$, such that the sum $\overline{v_z B_R} - \overline{v_R B_z}$ vanishes for the sunspot and is slightly negative for the naked spot consistent with the decay rate of about $10^{21}$ Mx~day$^{-1}$. In leading order the mixed polarity flux surrounding both spots is unrelated to sunspot decay, a decay manifests itself only in a small imbalance of these two terms, i.e. an analysis focused only on one of the terms is not meaningful.

We can also use Eq. (\ref{eq:flux_evol}) to estimate the potential contribution from numerical diffusivity to the decay of the naked spot. To this end we compare the flux evolution (left hand side) to the flux evolution estimated from the right hand side, which assumes ideal MHD without numerical diffusivity. We perform this analysis for the decaying naked spot from $t=50$ to $75$ hours. Since this simulation has been restarted from a previous simulation that was computed with a different formulation of numerical diffusivities and an additional flux correction term as described in Section \ref{Sec:setup}, it takes a few hours before it reaches a new equilibrium. In the first 3 hours (from $t=50-53$ hours) we find a difference between the left and right hand side of Eq. (\ref{eq:flux_evol}) of $10^{20}$ Mx/day, in the following $22$ hours (from $t=53-75$ hours) we find a difference of $5\times 10^{18}$ Mx/day, which is less than 
$1\%$ of the average decay rate of the spot. Overall this indicates that the direct contribution from numerical diffusivity is insignificant for the decay process. 

\begin{figure*}
  \centering 
  \resizebox{0.95\hsize}{!}{\includegraphics{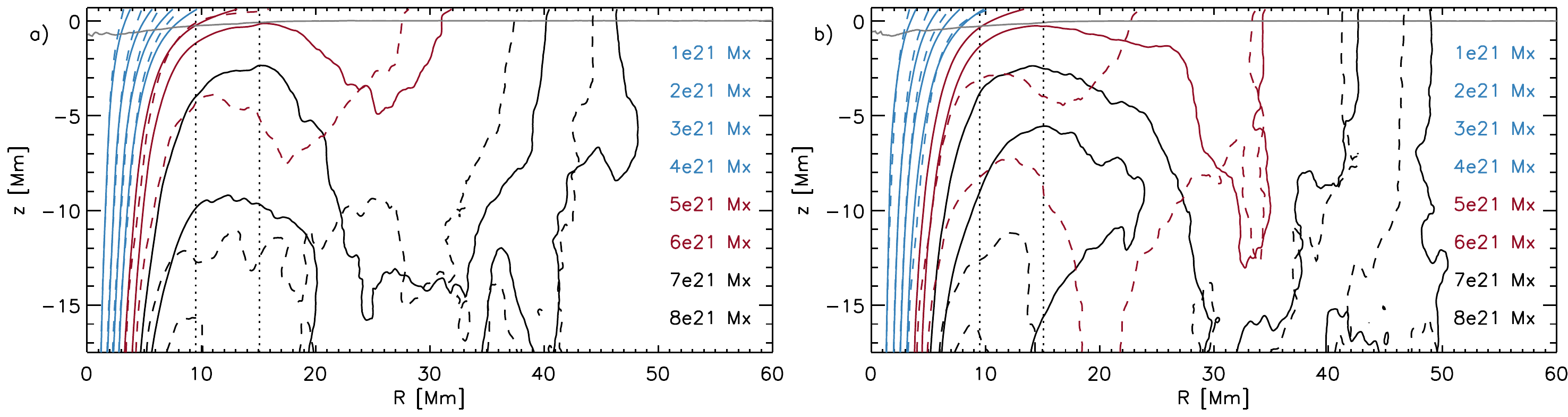}}
  \caption{Contour plots of the azimuthally averaged magnetic field structure for both spots. Flux surfaces of the sunspot are solid, flux surfaces of the naked spot are dashed. Panel a) shows the flux surfaces for a snapshot at $t=50$~h, panel b) at $t=75$~h. We show the flux surfaces in increments of $10^{21}$~Mx. Blue colors indicate flux surfaces corresponding to the umbra for both spots, red color shows flux surfaces that correspond to the penumbra in the case of the sunspot, dotted lines indicate the radius at which $\bar{I}=0.9 I_{\odot}$ for both spots. The grey line indicates the $\tau=1$ level.}
  \label{fig:08}
\end{figure*}

\subsection{Flow structure and thermal structure in proximity of spots}
In all following figures we will compare the same quantities for the sunspot and naked spot. The sunspot is always shown in the panels on the left and the naked spot in the panels on the right.

The connection between the large-scale time-averaged surface flow and subsurface flow is presented in Figure \ref{fig:06}. We show here flow maps that were averaged in time from $t=50 - 75$ hours and in addition horizontally smoothed through a convolution with a Gaussian with a FWHM of $3$~Mm. Panels a) and b) show the radial flow in the photosphere ($\tau=1$ level), panels c) and d) the radial flow in $4$~Mm depth and panels e) and f) the vertical flow in $4$~Mm depth. The most prominent difference between both spots is the direction of the flow in proximity of the spots in the photosphere. While the sunspot has an Evershed flow reaching an azimuthally averaged flow amplitude of $3-4$ km~s$^{-1}$ , the naked spot has an inflow of $1-2$ km~s$^{-1}$, which is due to the absence of a penumbra in the latter. Despite this difference both spots are surrounded further out by an outflow extending to about $1.5-2$ spot radii with amplitudes of a few 100 m~s$^{-1}$. The solid contour line encloses the region in which the outflow velocity is larger than $200$ m~s$^{-1}$. For both spots the overall extent of the radial surface flow region as well as non-axisymmetric shape is similar to that of the flow structure found in $4$~Mm depth. The corresponding vertical flow structure in $4$~Mm depth is such that for both spots an upflow occurs mostly beneath the outer boundary of the spots $I_{\odot}=0.9$ and an downflow occurs at the outer edge of the divergent flow region. In the case of the naked spot we see also some indication of a second ring of outflow in a distance of $30 - 40$~Mm from the spot center. This feature results from a ring-like arrangement of convection cells, which is also clearly visible in the azimuthal averages discussed later in Figures \ref{fig:09} and \ref{fig:10}. This feature is less pronounced for the sunspot. On a qualitative level we find similar flow systems around the sunspot and naked spot in $4$~Mm depth regardless of the difference in the photosphere. The flow systems around the sunspot are more extended and have an overall larger amplitude.

Figure \ref{fig:07} shows the thermal perturbations associated with the subsurface flow structure (the perturbation is relative to the mean temperature outside the $\bar{I}=0.9 I_{\odot}$ contour). Panels a) and b) show the  temperature perturbation averaged from $t=50 - 75$ hours and in addition spatially smoothed through a convolution with a Gaussian with a FWHM of $3$~Mm. Panels c) and d) show 15 minute averaged snapshots at $t=62.5$~hours (we use here a 15 minute average to remove the signature of p-modes). We find an average temperature increase of up to $20$~K in the region which also hosts the average subsurface upflow. The increase of the average temperature is due to the absence of cool downflows in this region. The temperature of individual convective upflows does not differ from that of a typical convective upflow outside the region influenced by the spots. While we do find a ``hot ring'' beneath the photosphere around both spots, we see no evidence for a ``bright ring'' in the photosphere (see also \citet{Rempel:2011:moat}). The brightness of the photosphere is determined by the entropy of upflow regions which remains unchanged.

\subsection{Comparison of the azimuthally averaged structure}

\begin{figure*}
  \centering 
  \resizebox{0.75\hsize}{!}{\includegraphics{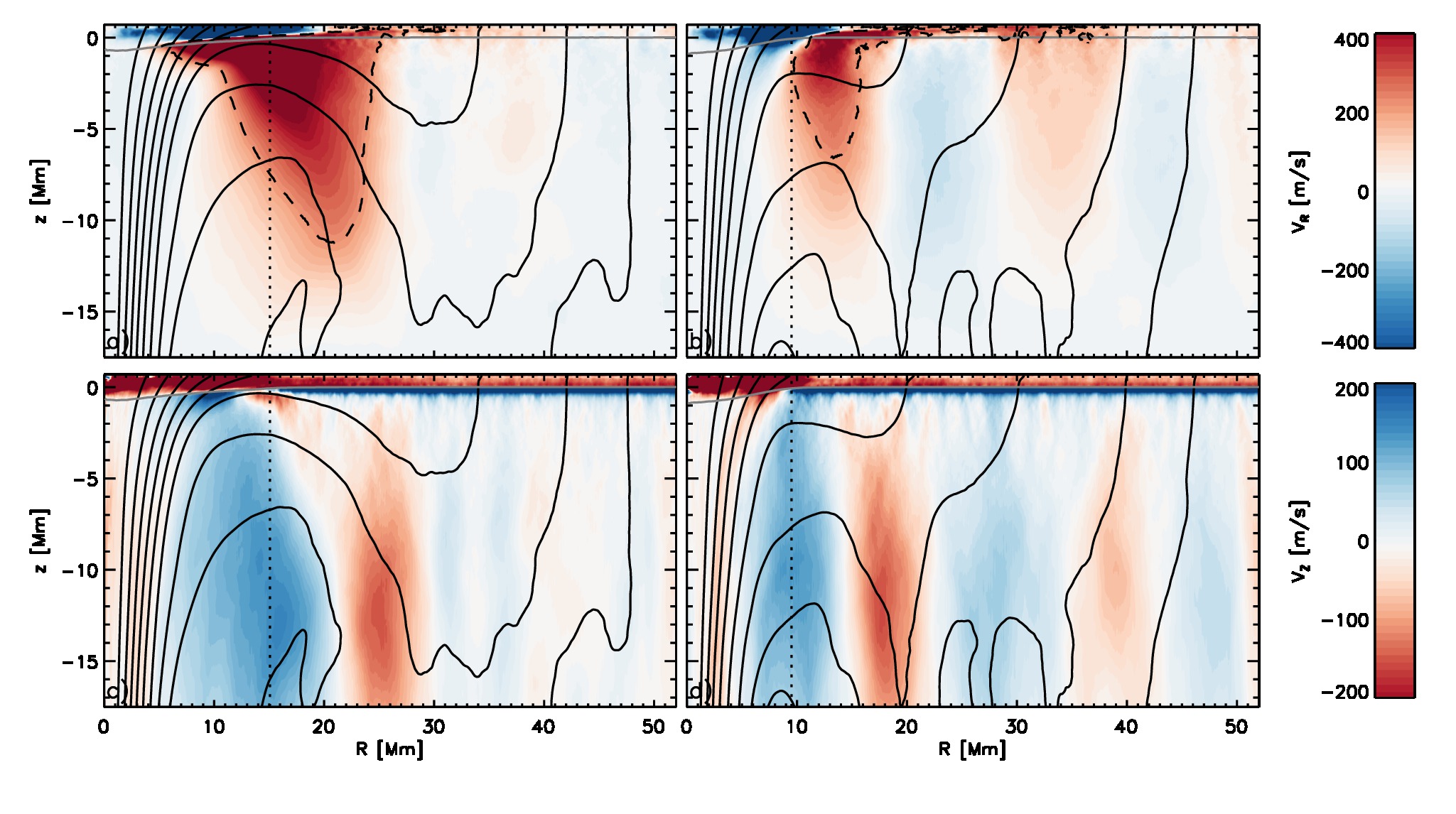}}
  \caption{Contour plots of the $25$~hour time averaged (from 50 to 75 hours) and azimuthally averaged radial (top) and vertical flow velocity (bottom). Positive values correspond to up/outflows. Black solid lines indicate the average flux surfaces, the grey solid line the $\tau=1$ surface, dotted lines indicate the radius at which $\bar{I}=0.9 I_{\odot}$. In addition dashed black contours enclose the region with radial outflows exceeding $200$~m~s$^{-1}$.} 
  \label{fig:09}
\end{figure*}

\begin{figure*}
  \centering 
  \resizebox{0.75\hsize}{!}{\includegraphics{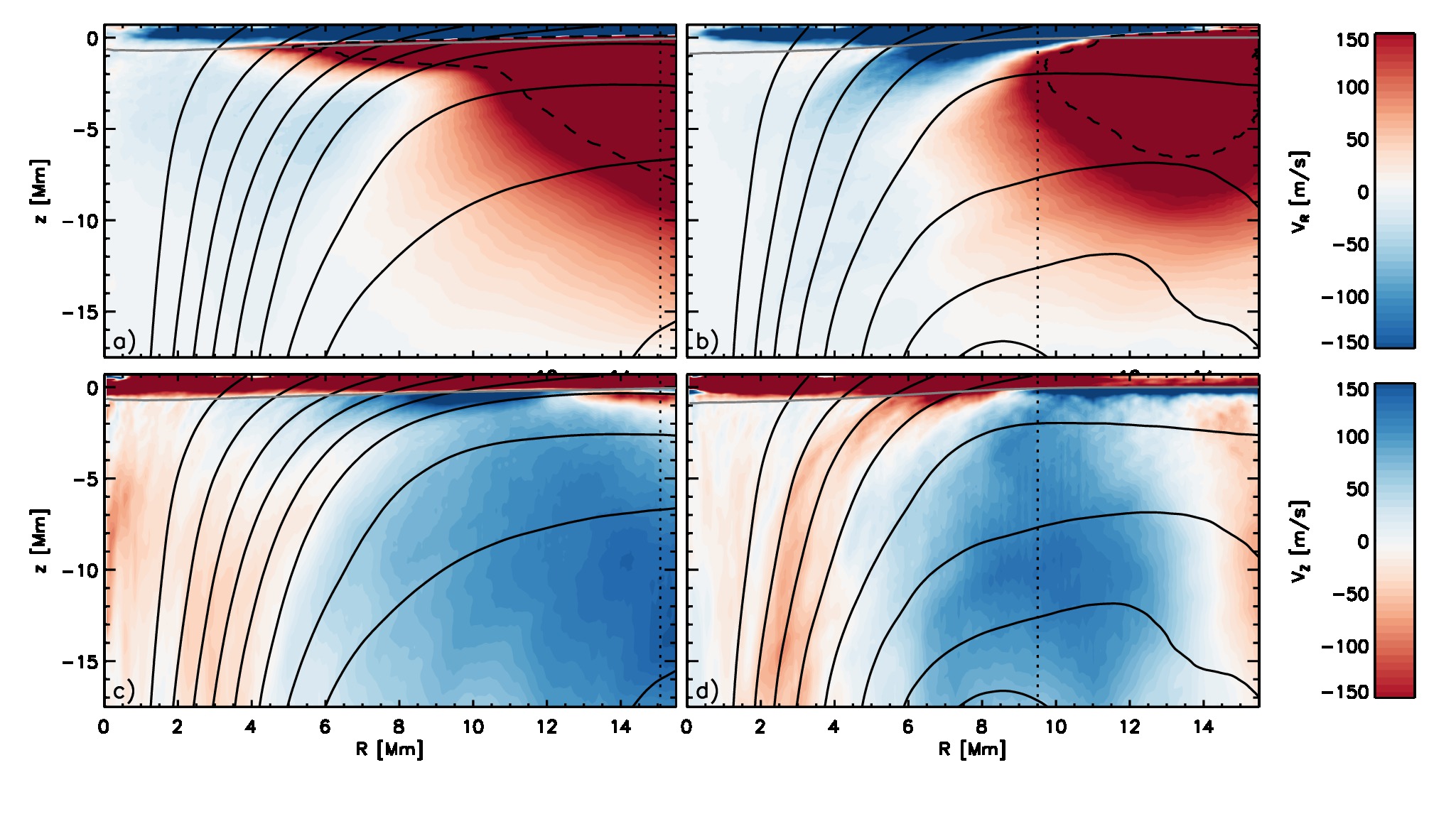}}
  \caption{Same as Figure \ref{fig:09}. The plot is focused on the center region $R<10$~ Mm in order to show the mean flows present underneath the spots.} 
  \label{fig:10}
\end{figure*}

\begin{figure*}
  \centering 
  \resizebox{0.75\hsize}{!}{\includegraphics{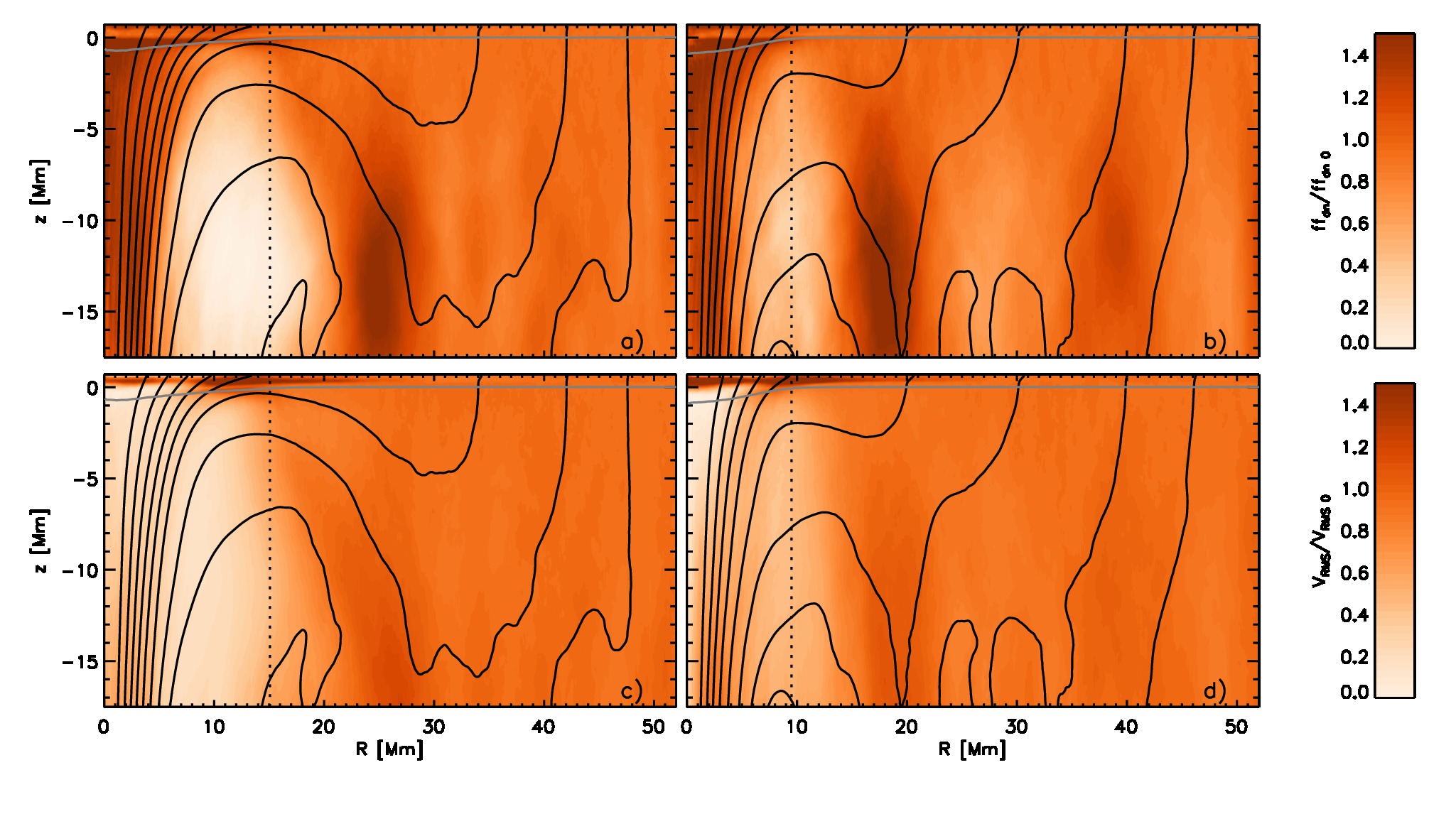}}
  \caption{Contour plots of the $25$~hour time averaged (from 50 to 75 hours) and azimuthally averaged convective RMS velocity (top) and downflow filling factor (bottom). Black solid lines indicate the average flux surfaces, the grey solid line the $\tau=1$ surface, dotted lines indicate the radius at which $\bar{I}=0.9 I_{\odot}$.}
  \label{fig:11}
\end{figure*}

In this section we compare both spots by looking in more detail at their azimuthally and temporally averaged mean structure (as before we  average from $t=50 - 75$ hours). Figure \ref{fig:08} compares the flux surfaces of both spots for a snapshot at $t=50$~h (panel a) and $t=75$~h (panel b). Solid lines correspond to the sunspot, dashed lines to the naked spot. We further color coded the flux surfaces. Blue colors correspond to the flux surfaces that are very similar for both spots. Red color are flux surfaces that correspond to the penumbra region in the case of the sunspot. At $t=50$~h the naked spot has a flux content of about $5\times 10^{21}$~Mx (see Figure \ref{fig:01}). Flux surfaces up to $5\times 10^{21}$~Mx are very similar, while the $6\times 10^{21}$~Mx flux surface is already submerged for the naked spot. At $t=75$~h also the $5\times 10^{21}$~Mx
flux surface becomes submerged for the naked spot, while there is no significant change for the sunspot (for the innermost $20$~Mm). Overall this points toward a mode of decay for the naked spot through submergence of horizontal magnetic field surrounding the spot. 

In order to understand better the differences in stability of both spots we compare in Figures \ref{fig:09} and  \ref{fig:10} the azimuthally and temporally mean flows. The top panels show radial, the bottom panels vertical mean velocity (averaged azimuthally and $25$~hours in time). Contour lines indicate the time averaged flux surfaces similar to Figure \ref{fig:08}. On a qualitative level both spots are surrounded by similar mean flows. The dominant subsurface flow is a diverging flow extending up to about 2 spot radii  (the photospheric spot radius based on the $\bar{I}=0.9 I_{\odot}$ contours is indicated by the dotted vertical lines). The corresponding vertical upflow peaks at a radial distance similar to the spot radius in both cases. The flow system extends to the bottom boundary of the simulation. While the vertical flow velocity is largest in the deeper parts of the domain, the radial flow amplitude is largest near the photosphere. Outflow velocities exceeding $200$~m~s$^{-1}$ are found to a depth of $11$~Mm for the sunspot and $7$~Mm for the naked spot. The only qualitative difference occurs right in the photosphere. While the sunspot has a strong $3-4$~km~s$^{-1}$ Evershed outflow, the naked spot is surrounded by a region with ~$1-2$~km~s$^{-1}$ inflows. Above the photosphere both spots show strong inflows. In addition there is also a very shallow outflow found above the $\tau=1$ level in a distance of more than two spot radii. This feature is most pronounced for the naked spot and has no relation to the direction of the subsurface flow. As discussed already in \citet{Rempel:2011:moat} this is caused by overshooting convection hitting the inclined magnetic canopy surrounding the spots. Figure \ref{fig:10}  shows the same quantities as Figure \ref{fig:09}, but is focused on the innermost $15$~Mm in order to highlight flows underneath the spots. Underneath both spots we find a weak converging flow with average amplitudes of a few $10$~m~s$^{-1}$. The most striking feature is in the case of the naked spot a concentrated downflow in-between the $2$ and $4\times 10^{21}$~Mx flux surface. This downflow with amplitudes reaching $100$~m~s$^{-1}$ drains the mass associated with the converging flow found in the photosphere. Since we show here azimuthal averages this flow appears to be present within the spot, in individual snapshots (not shown here) this flow is present right at the edge of the azimuthally warped boundary of the spot.

Figure \ref{fig:11} presents the average downflow filling factor (top panels) and convective RMS velocity (bottom panels). We show for both quantities changes relative to the average values found in a distance of more than $30$~Mm from the spot center for each respective height level. Also here both spots are on a qualitative level very similar. In the center underneath the umbra we find an enhancement of the downflow filling factor, further out we see a strong reduction of the downflow filling factor extending beyond the spot radius, followed again by a region with enhanced downflow filling factor. The convective RMS velocity is suppressed approximately within one spot radius. The main difference between both spots is as before the amplitude of these effects. Both, the suppression of downflow filling factor and convective RMS velocity in proximity of the spots are more pronounced for the sunspot. In the case of the sunspot the downflow filling factor is reduced by as much as
a factor of $12$, while the reduction for the naked spot is a factor of about $3$. The strong reduction of downflows in a region surrounding both spots is also evident from the temperature perturbation shown in Figure \ref{fig:07}c, d. The differences in the reduction of the convective RMS velocity are less dramatic, we find a factor of $4$ reduction for the sunspot and about a factor of $2$ reduction for the naked spot. 
  
After comparing in detail the thermal structure, mean flows  and convective properties in the proximity of both spots we now return to the question of why the naked spot decays at a rate of about $10^{21}$~Mx~day$^{-1}$, while the sunspot is almost stationary for the duration of the simulation. By comparing the average flux surfaces we found that the main difference between both spots is the submergence of the $5$ and $6\times 10^{21}$~Mx surface in the case of the naked spot. This submergence happens in a region which is for both spots characterized by a mean upflow, a mean
radial outflow, a suppression of downflow filling factor and convective RMS velocity. In particular the presence of a mean upflow, reduction of downflow filling factor and convective RMS velocity are factors that are a hurdle for submergence of field for both spots. The main difference is that these effects are much more pronounced in the case of the sunspot, in particular the reduction of the average downflow filling factor by a factor of up to $12$ is very striking: submergence of magnetic field is impossible in the absence of convective downflows.

\section{Physical cause of the moat flow}
The key element for understanding the origin of the moat flow is the strong reduction of downflows in a region surrounding both spots. The reduction of downflow filling factor is the consequence of two effects: (1) The reduction of the surface brightness within the penumbra and the equivalent region for the naked spot reduces the amount of low entropy material that is formed in the photosphere; (2) The presence of strong horizontal flows in this region (Evershed flow for the sunspot, converging flow for the naked spot) leads to a preferential draining of the low entropy material away from the spot (sunspot) or in close proximity of the spot (naked spot), leaving a region around the spots with a very small downflow filling factor.  For the effect (2) the flow direction appears to be of secondary importance, although the effect is clearly more pronounced for the sunspot. This is likely due to the more than 2 times stronger Evershed flow speed and the larger radial extent of the region with an Evershed flow compared to the inflow around the naked spot.

The reduced fraction of downflows perturbs the vertical mass flux balance. This leads naturally to an average upflow that requires a large scale outflow for reasons of continuity, the observed moat flow is the photospheric component of that flow. The upflow cannot disappear since the horizontal pressure balance in the convection zone imposes in this region a pressure gradient similar to that found in other upflow regions. The deeper seated component of this flow may be associated with the ``collar flows" that are discussed in \citep[e.g.,][]{Meyer:etal:1974,Hurlburt:Rucklidge:2000}. In addition the absence of cool downflows leads in this region to an increase of the average temperature as well as pressure. The subsurface ``hot ring'' does not lead to a ``bright ring'' in the photosphere since the entropy of upflows, which determines the brightness of the photosphere, remains unchanged. This explanation of the moat flow and absence of a bright ring was qualitatively discussed already in \citet{Spruit:1997} (see sections 5.3 and 5.4 therein). While our numerical solutions look on average similar to the models presented by \citet{Nye:1988:moat}, the explanation is somewhat different: In their model the moat flow is driven by a surplus of gas pressure beneath the penumbra caused by the temperature rise due to heat flux blockage, in \citet{Spruit:1997} and our simulations it is simply the consequence of a perturbation of the upflow/downflow balance around the spots. There is no plasma element that gets heated in this process, only the average temperature rises due to the reduction of contributions from cooler downflows. 

\citet{Rempel:2011:moat} also presented simplified models in which the sunspot was replaced by a cone-shaped obstacle or a heat blanked preventing radiative loss in the photosphere. In both cases a flow system with similar amplitude and extent was found. The flow systems we find in these simulations are the general response of the convection zone to an obstacle that impedes the formation of cool downflows in certain regions of the convection zone.

\section{Discussion and conclusion}
We presented two spot simulation setups that differ only with respect to the presence of a penumbra. We compared the properties of a simulated sunspot and naked spot with comparable size and analyzed how the presence of a penumbra influences the moat flow and sunspot decay.

\subsection{Moat flows}
With the exception of the Evershed flow, which is as expected only present when there is a penumbra, all other large scale flows surrounding the spots are qualitatively similar. The dominant flow pattern around the spots is a diverging flow that extends up to about 2 spot radii in agreement with \citet{Brickhouse:Labonte:1988}. The flow system is more extended for the spot with penumbra, since also the spot radius is larger. The outflow is present for both spots all the way to the bottom boundary of the simulation domain, although the flow amplitude is steadily declining at a rate comparable to that of the convective RMS velocity outside the spots. These findings confirm the earlier results reported by \citet{Rempel:2011:moat} and strongly suggest that moat and Evershed flow have a mostly independent origin. While the Evershed flow is a magnetized flow that results from overturning convection in a strongly inclined magnetic field, the moat flow is simply the result of perturbing the upflow/downflow balance in the proximity of a spot. The perturbation of the upflow/downflow balance around the spots leads to an unbalanced vertical mass flux that requires a radially overturning flow of which the photospheric component is the observed moat flow.   

Observationally it has not been settled whether moat-like outflows are limited to sunspots with penumbra or if they are also present around naked spots or even pores.  Observations of naked spots and pores \citep{Sobotka:etal:1999,VDominguez:etal:2010,SainzDalda:2012:naked} do show an inflow towards the spot in close proximity similar to the flow patterns shown in Figure \ref{fig:06}. However, they also show divergent flows further out similar to our findings. These outflows are typically not classified as ``moat flows'' and described as ``outward flows originating in the regular mesh of divergence centers around the pore'' \citep{VDominguez:etal:2010}. Our simulation results suggest that these flows are essentially identical to the moat flows found around sunspots. 

Already \citet{Brickhouse:Labonte:1988} pointed out that there is no significant correlation between the average moat properties (velocity and extent) and sunspot properties (size, age, stage of development). Similar results were found recently by \citet{Loehner-Boettcher:2013:moat}. Their analysis of the moat flow properties of 31 sunspots concluded that the moat flow velocity and moat region extent are not correlated with the sunspot size and Evershed flow speed. \citet{Sobotka:Roudier:2007,Loehner-Boettcher:2013:moat} found that the moat region extent is not simply proportional to the sunspot size. More work is required to determine if our simulation results are consistent with that, since we have at this point only one sunspot and one naked spot, while the observational result is based on sunspots with different sizes. However, we do find that the sunspot and naked spot have roughly the same photospheric flow profile with distance from the outer radius of the spot in each case, which implies that the moat extent is not simply proportional to the spot radius. \citet{Sobotka:Roudier:2007} (see their Equations 1 and 2) suggested a linear relation of the form $R_{\rm moat}=1.1 R_{\rm spot} + R_0$, where $R_0=7 (11)$~Mm for young (old) spots, which is qualitatively similar to Equation \ref{EQ:Rmoat}.

Our simulations show moat flows with a significant depth extent. For the sunspot setup the moat flow velocity exceeds $200$~m~s$^{-1}$ down to a depth of 
$11$~Mm, for the naked spot down to $7$~Mm. A similar depth extent is found in helioseismic inversions \citep{Featherstone:etal:2011:JPC}, although we do not find an indication of a secondary flow peak in about $5$~Mm depth as they do. One of the sunspot simulations discussed here was also recently analyzed by \citet{DeGrave:2014:sunspot} using time-distance helioseismology (the high resolution model discussed in their paper is the simulation that we ran for a total of 100 hours). It was found that the horizontal flow structure around the sunspots is recoverable in the uppermost $3-5$~Mm of the convection zone using a $25$ hour data set, although flow amplitudes are usually underestimated by up to $50\%$ compared to the flows present in the simulation. This might indicate that an in-depth comparison of our simulation results with flows from helioseismic inversions is only meaningful if the helioseismic inversion procedure is first tested on simulation data in order to quantify which aspects of the flow structure are recoverable.

An interesting feature we find for the naked spot simulation is the presence of radial mean flows that extend beyond the moat flow region. This pattern is due to a ring-like arrangement of convection cells, which leads on average to alternating converging and diverging flows around the spot, the moat flow is essentially the first cell of that pattern. This pattern is less pronounced for the sunspot, which might indicate sensitivity to the spot size. \citet{Svanda:2014:moat:SG} found such a feature in an average sense after determining the average moat flow from a sample of $104$ sunspots. They interpreted this feature as the contribution from neighboring supergranules in a distance of more than $20$~Mm from their average sunspot. 

\subsection{Sunspot decay}
We found two features that suppress sunspot decay: A strong reduction of the downflow filling factor and convective RMS velocity underneath the sunspot penumbra and the outer boundary of the naked spot. In particular the reduction of the downflow filling factor prevents the submergence of horizontal magnetic field, which turns out to be the dominant decay process in the simulations presented here. In the case of the sunspot these effects are so pronounced that we found a close to stationary solution starting from about $t=20$ to $100$ hours. In contrast to this the naked spot shows a decay at a rate of about $10^{21}$ Mx day$^{-1}$. We also do find a region with suppressed downflow filling factor and reduced RMS velocity around the naked spot, but these features are less pronounced compared to the sunspot. In several previous investigations \citep[see, e.g.,][]{Hurlburt:Rucklidge:2000,Botha:etal:2006} it has been suggested that a converging ``collar flow'' is required to stabilize sunspots against decay. We do not see here convincing evidence for that. While there is a weak converging subsurface flow beneath the umbra of both spots (Figure \ref{fig:10}), we do find in the case of the naked spot that enhancing that flow leads overall to a less stable configuration. 

We note that sunspot decay can result in addition from deeper seated convective motions that erode the ``footpoint'' of the spot and perhaps lead to flux separation an splitting of a spot. Such decay has been found to some degree in the simulations presented in \citet{Rempel:2011:moat,Rempel:Cheung:2014:fem}. Here we have chosen a setup that minimizes these effects in order to be able to study the decay caused by near surface flows.

\subsection{Moving magnetic features}
In observations the moat region of sunspots is also associated with so called moving magnetic features (MMFs) \citep{Harvey:Harvey:1973} that are very often linked to sunspot decay \citep[e.g.,][]{Martinez-Pillet:2002,Kubo:2008:mmf_decay,Kubo:2008:mmf_decay_err}. We do find around both spots a region with enhanced mixed polarity magnetic field with an extent of about $2 R_{\rm spot}$. The origin of the mixed polarity field is a magneto-convection process that is unrelated to spot decay, leading to large contributions of the $\overline{v_z B_R}$ and $\overline{v_R B_z}$ terms such that the total $\overline{v_z B_R-v_R B_z}$ remains close to zero. If there is spot decay, only a small imbalance (of the order of a few \%) of these two contributions is needed. At least in the simulations presented here an interpretation of a single component is physically not meaningful \citep[see also][for further discussion]{Rempel:Cheung:2014:fem}. The strength of the mixed polarity field is mostly dependent on the strength of the magnetic canopy, i.e. we find more mixed polarity flux around the sunspot. The extent of the region with enhanced mixed polarity magnetic flux differs from the extent of the region with outflows and the dependence of both on the spot radius has a different functional form. Observations show that moving magnetic features are also present around naked  spots \citep{Zuccarello:etal:2009} and it has been argued by \citet{SainzDalda:2012:naked} that this is due to a similar magnetic canopy structure surrounding sunspots and naked spots. We certainly find in our simulations that the strength of the magnetic canopy is the primary factor that determines the amount of mixed polarity field found around spots.

\subsection{Conclusion}
Overall we find that our simulated sunspot and naked spot are very similar with respect to large scale flow systems outside the spots. For both spots we find a photospheric moat flow extending about $10$~Mm beyond the spot boundary. This flow is the photospheric component of a deeper reaching outflow cell that is the consequence of a strong reduction of downflows underneath the sunspot penumbra and equivalent region for the naked spot. The resulting vertical mass flux imbalance requires a radially overturning large scale flow.  While for both spots the decay is inhibited by a strong reduction of the downflow filling factor and convective RMS velocity underneath the sunspot penumbra and the outer boundary of the naked spot, these effects are significantly more pronounced for the sunspot. As a consequence the sunspot turns out to be more stable (stationary flux content for the duration of the simulations) than the naked spot, which shows a steady decay of about $10^{21}$~Mx~day$^{-1}$. The presence of enhanced mixed polarity magnetic field in the moat region is in leading order the consequence of a magneto-convection process that is unrelated to spot decay, but strongly influenced by the strength of the magnetic canopy overlying the photosphere.

\acknowledgements
The National Center for Atmospheric Research (NCAR) is sponsored by the National Science Foundation. The author thanks Alberto Sainz Dalda and the anonymous referee for very helpful comments on the manuscript. We would like to acknowledge high-performance computing support from Yellowstone (http://n2t.net/ark:/85065/d7wd3xhc) provided by NCAR's Computational and Information Systems Laboratory, sponsored by the National Science Foundation, under project NHAO0002 and from the NASA High-End Computing (HEC) Program through the NASA Advanced Supercomputing (NAS) Division at Ames Research Center under project s9025. This research has been partially supported through NASA contracts NNH09AK02I (SDO Science Center) and NNH12CF68C.

\bibliographystyle{natbib/apj}
\bibliography{natbib/papref_m}

\begin{thebibliography}{47}
\expandafter\ifx\csname natexlab\endcsname\relax\def\natexlab#1{#1}\fi

\bibitem[{{Botha} {et~al.}(2006){Botha}, {Rucklidge}, \&
  {Hurlburt}}]{Botha:etal:2006}
{Botha}, G.~J.~J., {Rucklidge}, A.~M., \& {Hurlburt}, N.~E. 2006, \mnras, 369,
  1611, 1611

\bibitem[{{Botha} {et~al.}(2011){Botha}, {Rucklidge}, \&
  {Hurlburt}}]{Botha:etal:2011}
---. 2011, \apj, 731, 108, 108

\bibitem[{{Brickhouse} \& {Labonte}(1988)}]{Brickhouse:Labonte:1988}
{Brickhouse}, N.~S., \& {Labonte}, B.~J. 1988, \solphys, 115, 43, 43

\bibitem[{{Cheung} {et~al.}(2010){Cheung}, {Rempel}, {Title}, \&
  {Sch{\"u}ssler}}]{Cheung:etal:2010}
{Cheung}, M.~C.~M., {Rempel}, M., {Title}, A.~M., \& {Sch{\"u}ssler}, M. 2010,
  \apj, 720, 233, 233

\bibitem[{{Dedner} {et~al.}(2002){Dedner}, {Kemm}, {Kr{\"o}ner}, {Munz},
  {Schnitzer}, \& {Wesenberg}}]{Dedner:etal:2002:divB}
{Dedner}, A., {Kemm}, F., {Kr{\"o}ner}, D., {et~al.} 2002, Journal of
  Computational Physics, 175, 645, 645

\bibitem[{{DeGrave} {et~al.}(2014){DeGrave}, {Jackiewicz}, \&
  {Rempel}}]{DeGrave:2014:sunspot}
{DeGrave}, K., {Jackiewicz}, J., \& {Rempel}, M. 2014, \apj, 794, 18, 18

\bibitem[{{Deng} {et~al.}(2007){Deng}, {Choudhary}, {Tritschler}, {Denker},
  {Liu}, \& {Wang}}]{Deng:etal:2007:flow_decaying_sunspot}
{Deng}, N., {Choudhary}, D.~P., {Tritschler}, A., {et~al.} 2007, \apj, 671,
  1013, 1013

\bibitem[{{Evershed}(1909)}]{Evershed:1909}
{Evershed}, J. 1909, \mnras, 69, 454, 454

\bibitem[{{Featherstone} {et~al.}(2011){Featherstone}, {Hindman}, \&
  {Thompson}}]{Featherstone:etal:2011:JPC}
{Featherstone}, N.~A., {Hindman}, B.~W., \& {Thompson}, M.~J. 2011, Journal of
  Physics Conference Series, 271, 012002, 012002

\bibitem[{{Gizon} {et~al.}(2000){Gizon}, {Duvall}, \&
  {Larsen}}]{Gizon:etal:2000}
{Gizon}, L., {Duvall}, Jr., T.~L., \& {Larsen}, R.~M. 2000, Journal of
  Astrophysics and Astronomy, 21, 339, 339

\bibitem[{{Harvey} \& {Harvey}(1973)}]{Harvey:Harvey:1973}
{Harvey}, K., \& {Harvey}, J. 1973, \solphys, 28, 61, 61

\bibitem[{{Heinemann} {et~al.}(2007){Heinemann}, {Nordlund}, {Scharmer}, \&
  {Spruit}}]{Heinemann:etal:2007}
{Heinemann}, T., {Nordlund}, {\AA}., {Scharmer}, G.~B., \& {Spruit}, H.~C.
  2007, \apj, 669, 1390, 1390

\bibitem[{{Hurlburt} \& {Rucklidge}(2000)}]{Hurlburt:Rucklidge:2000}
{Hurlburt}, N.~E., \& {Rucklidge}, A.~M. 2000, \mnras, 314, 793, 793

\bibitem[{{Kitiashvili} {et~al.}(2009){Kitiashvili}, {Kosovichev}, {Wray}, \&
  {Mansour}}]{Kitiashvili:etal:2009}
{Kitiashvili}, I.~N., {Kosovichev}, A.~G., {Wray}, A.~A., \& {Mansour}, N.~N.
  2009, \apjl, 700, L178, L178

\bibitem[{{Kleint} \& {Sainz
  Dalda}(2013)}]{Kleint:Saina-Dalda:2013:unusual-filaments}
{Kleint}, L., \& {Sainz Dalda}, A. 2013, \apj, 770, 74, 74

\bibitem[{{Kubo} {et~al.}(2008{\natexlab{a}}){Kubo}, {Lites}, {Shimizu}, \&
  {Ichimoto}}]{Kubo:2008:mmf_decay_err}
{Kubo}, M., {Lites}, B.~W., {Shimizu}, T., \& {Ichimoto}, K.
  2008{\natexlab{a}}, \apj, 689, 1456, 1456

\bibitem[{{Kubo} {et~al.}(2008{\natexlab{b}}){Kubo}, {Lites}, {Shimizu}, \&
  {Ichimoto}}]{Kubo:2008:mmf_decay}
---. 2008{\natexlab{b}}, \apj, 686, 1447, 1447

\bibitem[{{L{\"o}hner-B{\"o}ttcher} \&
  {Schlichenmaier}(2013)}]{Loehner-Boettcher:2013:moat}
{L{\"o}hner-B{\"o}ttcher}, J., \& {Schlichenmaier}, R. 2013, \aap, 551, A105,
  A105

\bibitem[{{Louis} {et~al.}(2014){Louis}, {Beck}, {Mathew}, \&
  {Venkatakrishnan}}]{Louis:2014:anomalous_flow_penumbra}
{Louis}, R.~E., {Beck}, C., {Mathew}, S.~K., \& {Venkatakrishnan}, P. 2014,
  \aap, 570, A92, A92

\bibitem[{{Mart{\'{\i}}nez Pillet}(2002)}]{Martinez-Pillet:2002}
{Mart{\'{\i}}nez Pillet}, V. 2002, Astronomische Nachrichten, 323, 342, 342

\bibitem[{{Meyer} {et~al.}(1974){Meyer}, {Schmidt}, {Wilson}, \&
  {Weiss}}]{Meyer:etal:1974}
{Meyer}, F., {Schmidt}, H.~U., {Wilson}, P.~R., \& {Weiss}, N.~O. 1974, \mnras,
  169, 35, 35

\bibitem[{{Nye} {et~al.}(1988){Nye}, {Bruning}, \& {Labonte}}]{Nye:1988:moat}
{Nye}, A., {Bruning}, D., \& {Labonte}, B.~J. 1988, \solphys, 115, 251, 251

\bibitem[{{Parker}(1979)}]{Parker:1979:cluster}
{Parker}, E.~N. 1979, \apj, 230, 905, 905

\bibitem[{{Rempel}(2011{\natexlab{a}})}]{Rempel:2011}
{Rempel}, M. 2011{\natexlab{a}}, \apj, 729, 5, 5

\bibitem[{{Rempel}(2011{\natexlab{b}})}]{Rempel:2011:moat}
---. 2011{\natexlab{b}}, \apj, 740, 15, 15

\bibitem[{{Rempel}(2012)}]{Rempel:2012:penumbra}
---. 2012, \apj, 750, 62, 62

\bibitem[{{Rempel}(2014)}]{Rempel:2014:SSD}
---. 2014, \apj, 789, 132, 132

\bibitem[{{Rempel} \& {Cheung}(2014)}]{Rempel:Cheung:2014:fem}
{Rempel}, M., \& {Cheung}, M.~C.~M. 2014, \apj, 785, 90, 90

\bibitem[{{Rempel} {et~al.}(2009{\natexlab{a}}){Rempel}, {Sch{\"u}ssler},
  {Cameron}, \& {Kn{\"o}lker}}]{Rempel:etal:Science}
{Rempel}, M., {Sch{\"u}ssler}, M., {Cameron}, R.~H., \& {Kn{\"o}lker}, M.
  2009{\natexlab{a}}, Science, 325, 171, 171

\bibitem[{{Rempel} {et~al.}(2009{\natexlab{b}}){Rempel}, {Sch{\"u}ssler}, \&
  {Kn{\"o}lker}}]{Rempel:etal:2009}
{Rempel}, M., {Sch{\"u}ssler}, M., \& {Kn{\"o}lker}, M. 2009{\natexlab{b}},
  \apj, 691, 640, 640

\bibitem[{{Sainz Dalda} \& {Mart{\'{\i}}nez
  Pillet}(2005)}]{SDalda:MPillet:2005}
{Sainz Dalda}, A., \& {Mart{\'{\i}}nez Pillet}, V. 2005, \apj, 632, 1176, 1176

\bibitem[{{Sainz Dalda} {et~al.}(2012){Sainz Dalda}, {Vargas Dom{\'{\i}}nguez},
  \& {Tarbell}}]{SainzDalda:2012:naked}
{Sainz Dalda}, A., {Vargas Dom{\'{\i}}nguez}, S., \& {Tarbell}, T.~D. 2012,
  \apjl, 746, L13, L13

\bibitem[{{Scharmer} {et~al.}(2008){Scharmer}, {Nordlund}, \&
  {Heinemann}}]{Scharmer:etal:2008}
{Scharmer}, G.~B., {Nordlund}, {\AA}., \& {Heinemann}, T. 2008, \apjl, 677,
  L149, L149

\bibitem[{{Sheeley}(1969)}]{Sheeley:1969}
{Sheeley}, Jr., N.~R. 1969, \solphys, 9, 347, 347

\bibitem[{{Sheeley}(1972)}]{Sheeley:1972}
---. 1972, \solphys, 25, 98, 98

\bibitem[{{Sobotka} \& {Roudier}(2007)}]{Sobotka:Roudier:2007}
{Sobotka}, M., \& {Roudier}, T. 2007, \aap, 472, 277, 277

\bibitem[{{Sobotka} {et~al.}(1999){Sobotka}, {V{\' a}zquez}, {Bonet},
  {Hanslmeier}, \& {Hirzberger}}]{Sobotka:etal:1999}
{Sobotka}, M., {V{\' a}zquez}, M., {Bonet}, J.~A., {Hanslmeier}, A., \&
  {Hirzberger}, J. 1999, \apj, 511, 436, 436

\bibitem[{{Solanki}(2003)}]{Solanki:2003}
{Solanki}, S.~K. 2003, {\aap}r, 11, 153, 153

\bibitem[{{Spruit}(1997)}]{Spruit:1997}
{Spruit}, H. 1997, Memorie della Societa Astronomica Italiana, 68, 397, 397

\bibitem[{{V{\" o}gler} {et~al.}(2005){V{\" o}gler}, {Shelyag}, {Sch{\"
  u}ssler}, {Cattaneo}, {Emonet}, \& {Linde}}]{Voegler:etal:2005}
{V{\" o}gler}, A., {Shelyag}, S., {Sch{\" u}ssler}, M., {et~al.} 2005, \aap,
  429, 335, 335

\bibitem[{{{\v S}vanda} {et~al.}(2014){{\v S}vanda}, {Sobotka}, \&
  {B{\'a}rta}}]{Svanda:2014:moat:SG}
{{\v S}vanda}, M., {Sobotka}, M., \& {B{\'a}rta}, T. 2014, \apj, 790, 135, 135

\bibitem[{{Vargas Dom{\'{\i}}nguez} {et~al.}(2007){Vargas Dom{\'{\i}}nguez},
  {Bonet}, {Mart{\'{\i}}nez Pillet}, {Katsukawa}, {Kitakoshi}, \& {Rouppe van
  der Voort}}]{VDominguez:etal:2007}
{Vargas Dom{\'{\i}}nguez}, S., {Bonet}, J.~A., {Mart{\'{\i}}nez Pillet}, V.,
  {et~al.} 2007, \apjl, 660, L165, L165

\bibitem[{{Vargas Dom{\'{\i}}nguez} {et~al.}(2010){Vargas Dom{\'{\i}}nguez},
  {de Vicente}, {Bonet}, \& {Mart{\'{\i}}nez Pillet}}]{VDominguez:etal:2010}
{Vargas Dom{\'{\i}}nguez}, S., {de Vicente}, A., {Bonet}, J.~A., \&
  {Mart{\'{\i}}nez Pillet}, V. 2010, \aap, 516, A91, A91

\bibitem[{{Vargas Dom{\'{\i}}nguez} {et~al.}(2008){Vargas Dom{\'{\i}}nguez},
  {Rouppe van der Voort}, {Bonet}, {Mart{\'{\i}}nez Pillet}, {Van Noort}, \&
  {Katsukawa}}]{VDominguez:etal:2008}
{Vargas Dom{\'{\i}}nguez}, S., {Rouppe van der Voort}, L., {Bonet}, J.~A.,
  {et~al.} 2008, \apj, 679, 900, 900

\bibitem[{{Vrabec}(1971)}]{Vrabec:1971}
{Vrabec}, D. 1971, in IAU Symposium, Vol.~43, Solar Magnetic Fields, ed.
  R.~{Howard}, 329

\bibitem[{{Vrabec}(1974)}]{Vrabec:1974}
{Vrabec}, D. 1974, in IAU Symposium, Vol.~56, Chromospheric Fine Structure, ed.
  R.~G. {Athay}, 201

\bibitem[{{Zuccarello} {et~al.}(2009){Zuccarello}, {Romano}, {Guglielmino},
  {Centrone}, {Criscuoli}, {Ermolli}, {Berrilli}, \& {Del
  Moro}}]{Zuccarello:etal:2009}
{Zuccarello}, F., {Romano}, P., {Guglielmino}, S.~L., {et~al.} 2009, \aap, 500,
  L5, L5

\end{thebibliography}

\end{document}